\documentclass[a4paper,11pt]{article}
\pdfoutput=1 % if your are submitting a pdflatex (i.e. if you have
\usepackage{jheppub} % for details on the use of the package, please
\usepackage[T1]{fontenc} % if needed
\usepackage[utf8]{inputenc}
\usepackage{amsmath}
\usepackage{float}
\usepackage{lineno}
\usepackage{xcolor}
\usepackage{subfig}
\usepackage{array}
\usepackage{url}
\usepackage{hyperref}

\usepackage[outdir=source/figures/]{epstopdf}
\usepackage{amssymb}
\usepackage{accents}

\usepackage{lineno}
\emailAdd{vcarretero@km3net.de}
\emailAdd{km3net-pc@km3net.de}

\abstract{
KM3NeT/ORCA is a water Cherenkov neutrino detector under construction and anchored at the bottom of the Mediterranean Sea. The detector is designed to study oscillations of atmospheric neutrinos and determine the neutrino mass ordering. This paper focuses on an initial configuration of ORCA, referred to as ORCA6, which comprises six out of the foreseen 115 detection units of photo-sensors. A high-purity neutrino sample was extracted, corresponding to an exposure of 433 kton-years. The sample of 5828 neutrino candidates is analysed following a binned log-likelihood method in the reconstructed energy and cosine of the zenith angle. The atmospheric oscillation parameters are measured to be $\sin^2\theta_{23}= 0.51^{+0.04}_{-0.05}$, and $ \Delta m^2_{31} = 2.18^{+0.25}_{-0.35}\times 10^{-3}~\mathrm{eV^2} \cup \{-2.25,-1.76\}\times 10^{-3}~\mathrm{eV^2}$ at 68\% CL. The inverted neutrino mass ordering hypothesis is disfavoured with a p-value of 0.25.
}

\begin{document}

\title{Measurement of neutrino oscillation parameters with the first six detection units of KM3NeT/ORCA}

\author[a]{S.~Aiello,}
\author[b,bb]{A.~Albert,}
\author[c]{A.\,R.~Alhebsi,}
\author[d]{M.~Alshamsi,}
\author[e]{S. Alves Garre,}
\author[g,f]{A. Ambrosone,}
\author[h]{F.~Ameli,}
\author[i]{M.~Andre,}
\author[j]{L.~Aphecetche,}
\author[k]{M. Ardid,}
% ...

\affiliation[a]{INFN, Sezione di Catania, (INFN-CT) Via Santa Sofia 64, Catania, 95123 Italy}
\affiliation[b]{Universit{\'e}~de~Strasbourg,~CNRS,~IPHC~UMR~7178,~F-67000~Strasbourg,~France}
\affiliation[c]{Khalifa University, Department of Physics, PO Box 127788, Abu Dhabi, 0000 United Arab Emirates}
\affiliation[d]{Aix~Marseille~Univ,~CNRS/IN2P3,~CPPM,~Marseille,~France}
\affiliation[e]{IFIC - Instituto de F{\'\i}sica Corpuscular (CSIC - Universitat de Val{\`e}ncia), c/Catedr{\'a}tico Jos{\'e} Beltr{\'a}n, 2, 46980 Paterna, Valencia, Spain}
\affiliation[f]{INFN, Sezione di Napoli, Complesso Universitario di Monte S. Angelo, Via Cintia ed. G, Napoli, 80126 Italy}
\affiliation[g]{Universit{\`a} di Napoli ``Federico II'', Dip. Scienze Fisiche ``E. Pancini'', Complesso Universitario di Monte S. Angelo, Via Cintia ed. G, Napoli, 80126 Italy}
\affiliation[h]{INFN, Sezione di Roma, Piazzale Aldo Moro 2, Roma, 00185 Italy}
\affiliation[i]{Universitat Polit{\`e}cnica de Catalunya, Laboratori d'Aplicacions Bioac{\'u}stiques, Centre Tecnol{\`o}gic de Vilanova i la Geltr{\'u}, Avda. Rambla Exposici{\'o}, s/n, Vilanova i la Geltr{\'u}, 08800 Spain}
\affiliation[j]{Subatech, IMT Atlantique, IN2P3-CNRS, Nantes Universit{\'e}, 4 rue Alfred Kastler - La Chantrerie, Nantes, BP 20722 44307 France}
\affiliation[k]{Universitat Polit{\`e}cnica de Val{\`e}ncia, Instituto de Investigaci{\'o}n para la Gesti{\'o}n Integrada de las Zonas Costeras, C/ Paranimf, 1, Gandia, 46730 Spain}
% ...
\affiliation[bb]{Universit{\'e} de Haute Alsace, rue des Fr{\`e}res Lumi{\`e}re, 68093 Mulhouse Cedex, France}
% ...
% ----- End address list
% ----- End automatically generated KM3NeT info

\maketitle
\flushbottom

\section{Introduction}
\label{sec:intro}

The discovery of neutrino oscillations and non-zero neutrino masses represents a deviation from the prediction of the Standard Model of particle physics \cite{RevModPhys.88.030501}. The theoretical framework for neutrino oscillations was first introduced after an initial intuition of Pontecorvo, by Maki, Nakagawa, and Sakata in 1962 \cite{Pontecorvo1957MesoniumAA, Maki:1962mu}. 
Since then, the three-flavour neutrino model with neutrino masses has become a well-established framework, with many oscillation parameters being measured with improving precision \cite{Esteban:2020cvm,deSalas:2020pgw, Capozzi:2021fjo}. However, several questions persist, including the value of the Dirac CP phase ($\delta$), the octant of the mixing angle $\theta_{23}$, ($< 45^\circ$ or $> 45^\circ$), and the neutrino mass ordering (NMO), which is either normal, NO, ($m_1 < m_2 \ll m_3$) or inverted, IO, ($m_3 \ll m_1 < m_2$). These questions are actively pursued in experimental efforts to enhance our understanding of neutrino properties.

Atmospheric neutrinos, generated by cosmic ray interactions in the Earth's atmosphere, are good probes for neutrino oscillation studies. They arrive from all directions with different propagation lengths ($L$) and they are emitted in a broad range of energies ($E$), hence probing a large range of $L/E$ values. This makes them particularly valuable for investigating oscillation phenomena. Large-volume Cherenkov detectors, such as the ORCA underwater neutrino telescope, offer the possibility to detect atmospheric neutrinos in the GeV-PeV range.

\begin{figure}[H]
    \centering
     \subfloat{
    \includegraphics[width=0.48\linewidth]{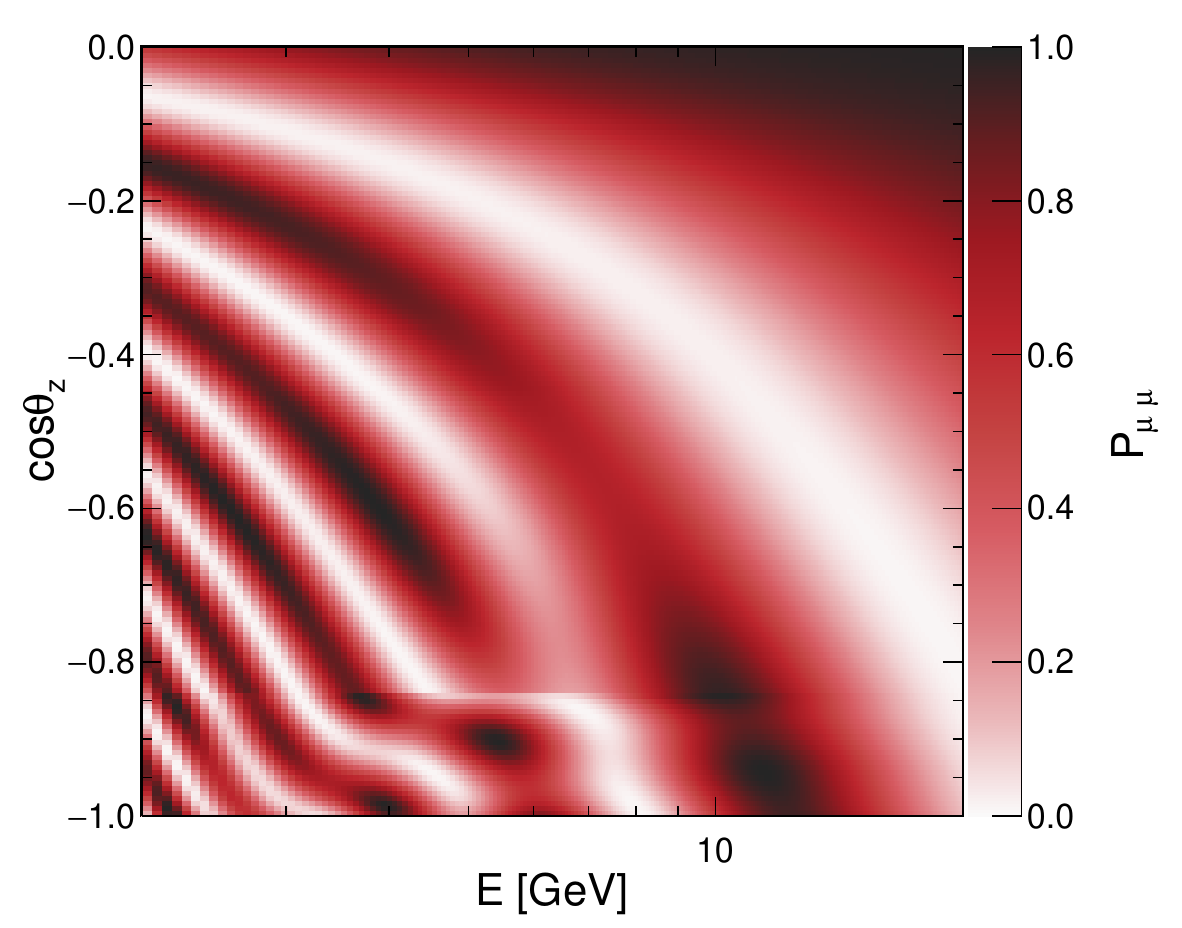}}
          \subfloat{   
      \includegraphics[width=0.48\linewidth]{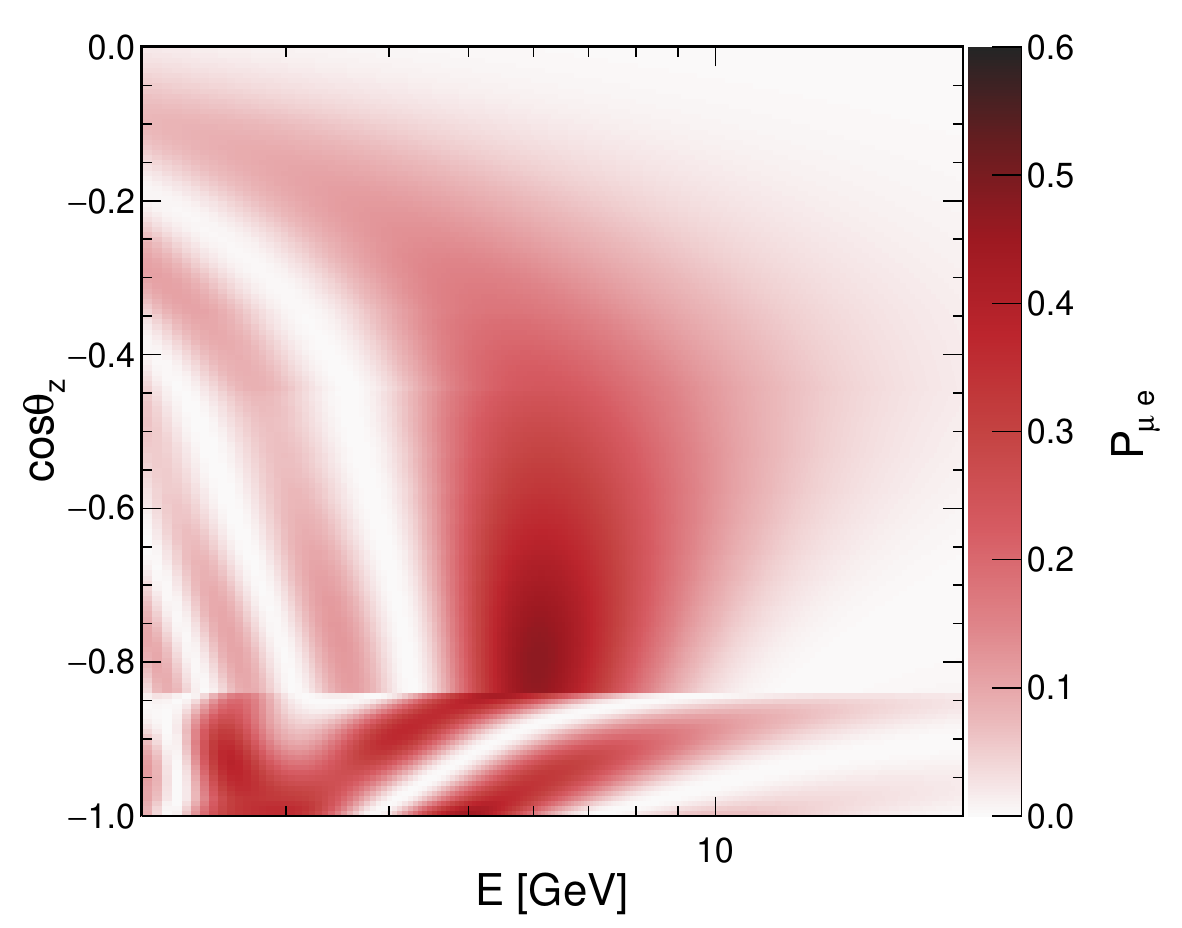}}
     \caption{The probabilities for muon neutrino survival, $P_{\mu\mu}$, (left) and transition to electron neutrino, $P_{\mu e}$, (right) for NO as a function of cosine of the zenith angle and neutrino energy. NuFit v5.2 best-fit parameters with atmospheric neutrino data from Super-Kamiokande are used \cite{Esteban:2020cvm}.}
    \label{F1:Osc}
\end{figure}

The atmospheric neutrino flux is dominated by muon neutrinos. The probabilities for muon neutrino survival and transition to electron neutrino due to oscillations are shown in figure~\ref{F1:Osc} as a function of the neutrino energy and of the cosine of the zenith angle (measured from the upwards vertical at the detector location), $\cos\theta$. The latter indicates the incoming neutrino direction from which the propagation length can be evaluated. The oscillation probabilities have been computed with OscProb \cite{joao_coelho_2023_10104847} using the Preliminary Reference Earth Model \cite{PREM}. The KM3NeT/ORCA detector is designed to measure neutrinos in the energy range of resonant matter effects, i.e. mainly from 5 to 10 GeV, where sensitivity to the NMO arises through both an enhancement of electron neutrino appearance and a distortion in the pattern of muon neutrino disappearance \cite{Adri_n_Mart_nez_2016}. The electron neutrino appearance signal also provides sensitivity to the octant of $\theta_{23}$. Neutrinos of higher energy provide information on the first minimum of the muon neutrino survival probability, which is sensitive to $\Delta m^2_{31}$ and $\sin^22\theta_{23}$. The pattern distortions near the vertical are due to matter effects and variations in the Earth density \cite{Smirnov:2003da}.

In this paper the first analysis of the complete dataset collected with an early sub-array of the KM3NeT/ORCA detector with six detection units, and called hereafter ORCA6 \cite{Adri_n_Mart_nez_2016}, is presented. The measurement of the $\Delta m^2_{31}$ and $\theta_{23}$ parameters that characterise atmospheric oscillations is detailed alongside a first probe of the neutrino mass ordering. Previously, studies have been conducted with a subset of this sample \cite{KM3NeT:2021hkj,Carretero:2022mjc}. The data sample presented in this paper contains 510 days of data collected from January 2020 to November 2021 for an exposure of 433 kton-years, corresponding to a 46\% increase in exposure with respect to previous results. The analysis encompasses improvements in simulation, reconstruction, and particle identification. 

The paper is organised as follows: in section~\ref{sec:detector}, the ORCA6 neutrino detector, technology, detection principle, and Monte Carlo (MC) simulations are described. In section~\ref{sec:datasample} details of the data taking period and data selection are reported.  In section~\ref{sec:analysis} the analysis methods are explained, and in section~\ref{sec:results} the results of the data analysis are presented.

\section{The KM3NeT/ORCA detector}
\label{sec:detector}

The KM3NeT Collaboration \cite{Adri_n_Mart_nez_2016} is building two neutrino detectors at the bottom of the Mediterranean Sea. The KM3NeT/ARCA detector (Astroparticle Research with Cosmics in the Abyss) is being built 100 km offshore the Sicilian coast near Portopalo di Capo Passero (Italy), anchored at a depth of 3500~m. Its design is optimised for the detection of high-energy neutrinos from astrophysical sources in the TeV$-$PeV range. The KM3NeT/ORCA detector (Oscillation Research with Cosmics in the Abyss) is being constructed close to the ANTARES site \cite{ANTARES:1999fhm} near the coast of Toulon (France), 40~km offshore at 2500~m depth. The KM3NeT/ORCA detector is optimised for the determination of the neutrino mass ordering \cite{NMOpaper, KM3NeT:2021rkn} and can probe physics beyond the Standard Model \cite{KM3NeT:2021uez,Decay_ORCA_2023, Domi:2023uy} in the neutrino sector using the atmospheric neutrino flux in the GeV$-$TeV range.

\subsection{Technology and layout}

The detectors consist of 3-dimensional arrays of photo-sensors  hosted in pressure-resistant glass spheres: the Digital Optical Modules (DOMs) \cite{KM3NeT:2022pnv}. Each DOM  houses 31 photomultiplier tubes (PMTs) with their associated readout electronics and calibration instrumentation. The PMTs are distributed over almost the full $4\pi$ solid angle, with more PMTs in the bottom hemisphere to enhance the detection of up-going particles. The DOMs are arranged along the Detection Units (DUs), flexible support structures which are anchored to the seabed and kept vertical by the buoyancy of the DOMs and by a submerged buoy at the top.  Each DU supports 18 DOMs by two parallel thin ropes; a backbone cable runs the full length of the DU. KM3NeT/ORCA will comprise 115 DUs, with an average horizontal spacing between the DUs $\sim20$~m; the vertical spacing between DOMs on a DU is $\sim9$~m. The final instrumented volume will be about  $7\times 10^6~\mathrm{m^3}$.

\begin{figure}[h]
    \centering
        \includegraphics[width=1\linewidth]{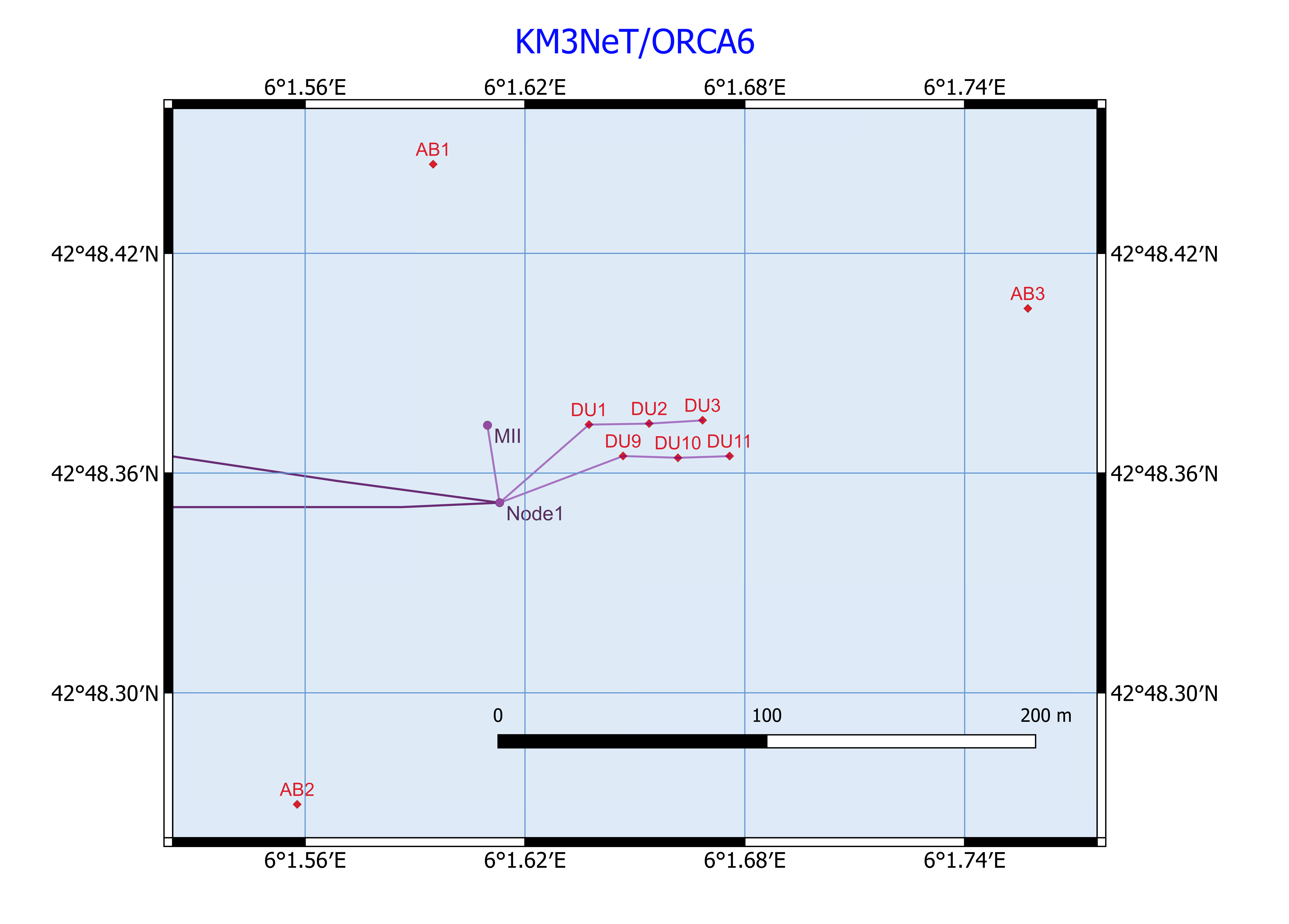}
        \caption{Layout of the ORCA6 detector configuration. The dots labelled as \textit{DU}n in the diagram represent detection
units, connected to the node. The node is connected
with the primary electro-optical cable, which connects the detector with the control station onshore. The dots marked as AB1-3 are autonomous acoustic beacons used for the position calibration of the detector. The Module Interface Instrumented (MII)  provides environmental data such as  temperature, conductivity/salinity and pressure \cite{75839}.}
    \label{fig:ORCA6_footprint}
\end{figure}

During the initial phase referred to as ORCA6, the detector was operating with six out of the total 115 DUs, with an  instrumented volume about $4\times 10^5~\mathrm{m^3}$. The footprint of the detector configuration is shown in figure~\ref{fig:ORCA6_footprint}.

\subsection{Detection and trigger}
\label{sec:trigger}
The detection principle is based on the observation of Cherenkov radiation induced by relativistic charged particles in the seawater. The Cherenkov photons propagate through seawater and reach the PMTs, generating a \textit{hit}. A hit contains the PMT identifier, the time of the photon arrival and the time during which the signal is over a certain threshold. Apart from the Cherenkov photons produced in neutrino interactions also background hits are expected in the data taking of KM3NeT/ORCA. These are due to the spontaneous electron emission of the PMTs, known as dark count \cite{Aiello_2018}, to the bioluminescence produced from a diverse range of living organisms in the deep sea \cite{annurev-marine-120308-081028} and to the natural radioactivity of $^{40}$K in the seawater \cite{Albert_2018}. The detector is installed at a depth of 2.5~km to provide shielding against atmospheric muons produced in cosmic ray interactions in the Earth's atmosphere. However, since high-energy muons can still reach the detector, event selections are needed to remove them.

The KM3NeT detectors employ online data filtering at the onshore data processing centre to select neutrino interaction events while suppressing the mentioned backgrounds.  The trigger conditions are based on causally related hits. In the  energy range relevant for KM3NeT/ORCA neutrino interactions in seawater primarily involve deep inelastic scattering with nucleons, although quasielastic scattering and resonant meson production cannot be disregarded \cite{Formaggio_2012}. Neutral current (NC) neutrino interactions produce a neutrino and a hadronic shower, while charged current (CC) neutrino interactions result in a hadronic shower and a charged lepton. The events in KM3NeT/ORCA are classified into two categories: "track-like" events encompassing $\nu_{\mu}$~CC events and $\nu_{\tau}$~CC events followed by the tau lepton decaying into a muon (branching ratio of about 17$\,\%$ \cite{PDG}); and "shower-like" events, comprising $\nu_{e}$~CC events, other decay channels of the tau lepton from $\nu_{\tau}$~CC events, and all flavour $\nu$ NC events. The same classification scheme applies to antineutrinos. This distinction is motivated by the different event topologies in the detector. 

Muons in the GeV range are minimum ionising particles and travel in a straight line until they stop. The muon track length is proportional to its energy with a range of approximately 4~m/GeV. For this reason, the 3D-muon trigger identifies causally connected hits within the volume of a hypothetical cylinder surrounding the active detector elements.

Electrons above some tens of MeV lose energy mainly through bremsstrahlung, and photons convert to $e^+e^-$ pairs, leading to electromagnetic showers. Both CC and NC interactions produce hadronic showers. The lifetime of the $\tau$ lepton is very short ($\tau_{\tau}=2.9\times 10^{-13}$~s) and at GeV energies the vertices of the neutrino interaction and of the tau lepton decay cannot be resolved in KM3NeT/ORCA. Two triggers are dedicated on shower events, the 3D-shower trigger, that identifies causally connected hits in the volume of a sphere of predefined radius and the MX-shower trigger, optimised for low-energy events, with a relaxed condition on the causal relation among the hits.

More details about the data acquisition are given in ref.~\cite{refId0}. Regardless of which algorithm triggered the events, they are reconstructed assuming both a track and a shower hypothesis. The reconstruction is performed using the probability density function (PDF) of the photon arrival times at the PMTs that are stored in lookup tables for each topology. More information on the track reconstruction is given in ref.~\cite{OFearraigh:2024ioy}; details about the shower reconstruction can be found in ref.~\cite{Alba}.

Time and position calibration procedures are crucial for ensuring an accurate event reconstruction \cite{KM3NeT:2021lsb, Bailly-Salins:2023IA, SanchezLosa:2023v3}. Intra-DOM, inter-DOM and inter-DU relative time calibration is achieved by calibration procedures before deployment of the DUs and \textit{in situ} calibrations using the background rates of $^{40}$K, atmospheric muons and nanobeacons. The position and orientation of the DOMs are continuously monitored using acoustic positioning systems and compasses \cite{GatiusOliver:2023qr}.

\subsection{Monte Carlo simulation}
\label{sec:sim}
The detector response is simulated with Monte Carlo events. Neutrino events in KM3NeT/ORCA are generated with gSeaGen \cite{Aiello_2020}, a GENIE-based \cite{andreopoulos2015genie} application developed to efficiently generate large samples of events induced by neutrino interactions and detectable in a neutrino telescope. Cherenkov photons induced by charged particles are propagated to the PMTs by the KM3NeT package KM3Sim \cite{Tsirigotis:2011zza}, based on Geant4 \cite{AGOSTINELLI2003250}, taking into account the light absorption and scattering in seawater. To speed up the simulation of light propagation in the case of high-energy particles, a custom KM3NeT package that uses precomputed tables of PDFs of the light arrival time is used. To generate atmospheric muon events MUPAGE \cite{Carminati:2008qb, Becherini:2005sr} simulates muon bundles at the detector surface. The optical backgrounds due to the PMT dark count rate and to the decay of $^{40}\mathrm{K}$ present in seawater are included through a KM3NeT package, which also simulates the digitised output of PMT responses and the readouts. From here onwards, both simulated and data events follow the same chain of triggering and reconstruction. More details about the simulations can be found in refs. \cite{Adri_n_Mart_nez_2016, NMOpaper}.

\section{Data sample}
\label{sec:datasample}
 \subsection{Run selection}

The ORCA6 detector was operational from January 2020 to November 2021. Data were taken continuously and divided into \emph{runs} with a typical duration of 6 hours. Data quality criteria were applied to exclude runs taken in unstable data-acquisition conditions, or with poor timing accuracy.

The KM3NeT detector readout system is based on the \emph{all-data-to-shore} concept, digitising all analog PMT signals over a preset threshold (typically 0.3 photoelectrons) and sent to shore for real-time processing. Physics events are filtered from the background using dedicated software which organises the continuum stream of data in intervals with a certain duration (timeslices); therefore a continuous data filtering is applied to keep consistency. Embedded in User Dataframe Protocols (UDP), timeslices from each DOM are fragmented in frames, associated with an absolute timestamp, and successively reconstructed onshore via a First In, First Out (FIFO) data buffer. In case of high-rate data transmission, some UDP packets can get lost compromising their acquisition onshore. Runs in which the UDP packet loss is above 5\% were rejected, as well as runs in which the FIFO was almost full.

Finally, the stored optical data contains the time and the time-over-threshold of each analog pulse. Additionally, summary data containing the rates of all PMTs in the detector are stored with a sampling frequency of 10 Hz. This information is used in the simulations and the event reconstruction to take into account the current status and optical background conditions of the detector. The dark count rate measured in the lab is about 1 kHz \cite{Aiello_2018}, while in seawater PMT rates are about 7 kHz due to $^{40}$K decays plus a variable contribution from bioluminescence. PMTs with rates higher than 20 kHz are not used in the reconstruction. Runs containing more than 50\% of unused PMTs on average are also excluded from the analysis.

After run selection, a total of 510 days out of the 633 days of data taking remain. The cumulative exposure of the KM3NeT/ORCA detector is defined assuming each active PMT instruments 108.8 tons of seawater. For the full detector (115 DUs), this would correspond to an instrumented mass of 6.98 Mton. Multiplying the number of active PMTs by the livetime for each run, the total exposure is determined. The run selection for the ORCA6 dataset resulted in an exposure of 433 kton-years.

\subsection{Event selection}

The data sample is largely dominated by the optical background. The multi-PMT structure of the KM3NeT DOM is exploited by online trigger algorithms in order to retain causally-correlated photon hits, as explained in section \ref{sec:trigger}. Sets of hits contributing to at least one of these triggers are called triggered hits. The single-PMT rate, however, is large enough to still give a non-negligible contribution of random coincidences surviving the trigger filtering and producing events. These events need to be removed from the dataset with the application of a base selection level (referred to as \textit{pre-cuts}):

\begin{enumerate}
\item The track reconstruction likelihood ratio between signal+background and only background hypotheses is required to be above 40. This parameter measures the agreement between the observed and expected arrival times of hits for the hypothesis of a muon track. Events triggered on optical background showed a value below 30.

\item At least 15 triggered hit per event are required to reliably remove all events which are exclusively due to optical background.

\end{enumerate}

The application of these selection criteria reduce the optical background to negligible levels. The optical background was not simulated for this analysis and its contribution was estimated using data.

\begin{figure}[H]
 \centering
\includegraphics[width=0.495\linewidth]{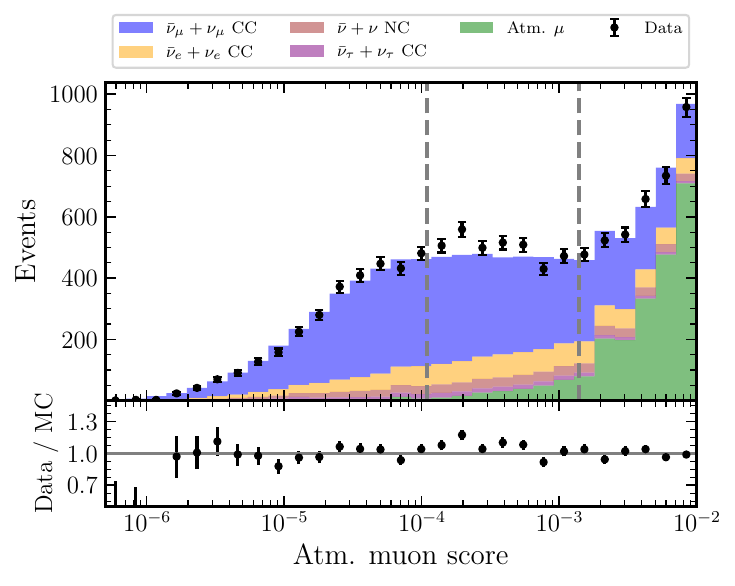} \includegraphics[width=0.495\linewidth]{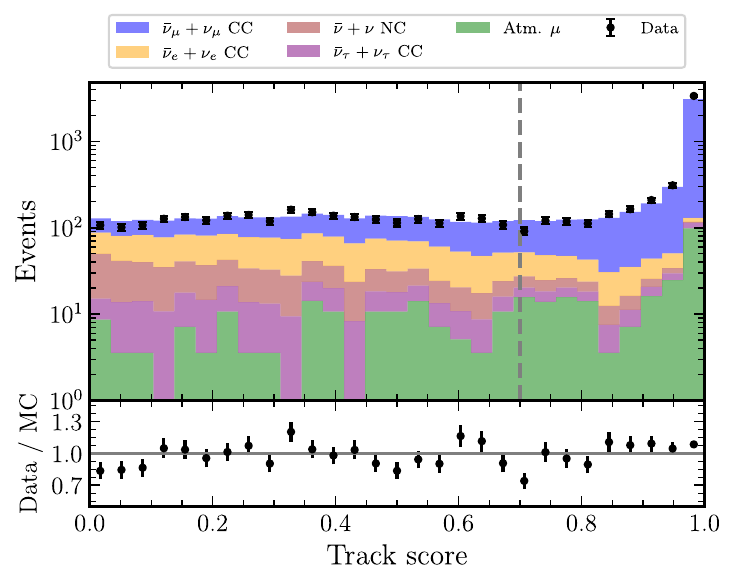}
    \caption{Atmospheric muon score, $\mu_s$ (left) and track score, $t_s$ (right) distributions for MC and data. Vertical lines mark the applied selection cuts. The track score distribution includes only events with an atmospheric muon score $\mu_s \leq 1.8\times 10^{-3}$.}
    \label{Fig:C6-PIDdata}
\end{figure}

After this first level of the event selection, the data sample consists of more than 99\% atmospheric muons. Neutrino candidates are selected by requiring a  track reconstructed as up-going. Occasionally, atmospheric muons can be reconstructed as up-going by the track reconstruction algorithm. Dedicated Boosted Decision Trees (BDTs) are trained to distinguish neutrino interactions and misreconstructed atmospheric muons. Using the XGBoost library \cite{Chen_2016}, individual scores are assigned to each event, interpreted as relating to the probability of the event belonging to either of the target classes (neutrino or muon). The BDT is trained on an unweighted sample of $10^{5}$ up-going MC events of each class which pass the pre-cuts, using 5-fold cross-validation and 150 decision trees, in order to apply the resulting model to the remaining dataset. 
Different variables based on the PMT hit distributions and the reconstruction outputs are used to build the nodes of the decision trees.
A set of 23 variables is chosen for their power in separating the target classes and for the agreement of their event distributions between data and MC. The most powerful ones are the mean vertical position of hits compatible with Cherenkov light emission (called "Cherenkov hits"), and the difference in the number of Cherenkov hits recorded in the upper and lower hemispheres of the DOMs. Atmospheric muons are expected to trigger mostly the upper part of the detector, thus having higher mean vertical position and more hits in the upper hemisphere than neutrino-induced muons. The resulting BDT score distribution, named atmospheric muon score $\mu_s$, is shown in figure~\ref{Fig:C6-PIDdata}; higher values are indicative of  muon-like events. By rejecting events with $\mu_s > 1.8 \times 10^{-3}$ the atmospheric muon background contamination of the sample is reduced to 2\%.

The last selection stage aims at isolating track-like neutrino interactions from the shower-like ones, since the different neutrino flavours serve to constrain complementary oscillation effects. A second BDT is trained using $10^{5}$ MC events of each class, 400 decision trees and 43 training variables, some of which coincide with those used for the muon score BDT. The more demanding settings compared to the previous BDT is a reflection of the difficulty of separating low-energy tracks from showers in a detector of small size. The extra variables used for this BDT make use of the difference in reconstruction output between the track and shower dedicated reconstruction algorithms for the same event; for instance, the angular deviation between the direction reconstructed by the track and by the shower algorithms is a powerful flavour discriminator, as $\nu_\mu$ CC interactions exhibit more compatibility between those two directions than $\nu_e$ interactions. Further efficient discriminators are based on the compatibility of the observed hit distribution with either track or shower geometry. The output of the BDT is the track score, $t_s$, where values closer to 1 are more compatible with a track-like topology as can be seen in figure~\ref{Fig:C6-PIDdata}. While $\nu_{\mu}$ and $\bar{\nu}_{\mu}$  events produce a strong peak at high track score values, indicating a clear identification of track-like topologies, the distribution for  $\nu_{e}$ and $\bar{\nu}_{e}$ events shows a more gradual increase towards lower scores without a pronounced peak near zero. This suggests that, although distinguishing track-like events is straightforward, accurately classifying showers as such remains challenging with the current detector size. The ability to identify showers is expected to improve significantly with a larger detector \cite{NMOpaper}.

The dataset is then categorised into three primary classes: "High Purity Tracks", "Low Purity Tracks", and "Showers". The distinction between tracks and showers is based on the track score, $t_s$, and the differentiation between High Purity Tracks and Low Purity Tracks is based on the atmospheric muon score, $\mu_s$. The atmospheric muon score correlates with the reconstruction quality and helps distinguish well-reconstructed tracks. The cuts in track score and atmospheric muon score are optimised to enhance sensitivity to the oscillation parameters $\Delta m^2_{31}$ and $\theta_{23}$:

\begin{itemize}
\item Events with an atmospheric muon score 
 $\mu_s > 1.8\times 10^{-3}$ are considered background and are discarded.
\item Events with a track score $t_s>0.7$ are classified as track-like, while remaining events are classified as shower-like.
\item Track-like events are separated into two classes of $\mu_s$. "High purity Tracks", events with a stricter selection $\mu_s<1.1\times 10^{-4}$, while other events are classified as "Low Purity Tracks".
\end{itemize}

 In this analysis, no events are simulated above 10~TeV in energy. A tail of high-energy $\nu_{\mu}$ CC events may be present in data. Therefore, events with reconstructed energies above 100~GeV were excluded to suppress the contribution of events that are not simulated to a negligible level. In contrast, shower events are mostly contained and the energy resolution is better, so that the reconstructed energy cut can be relaxed up to 1~TeV, allowing these events to be used for constraining nuisance parameters. The performance of the energy reconstruction is shown in figure~\ref{Fig:RecoE}. The track energy is essentially estimated from the measured track length. Due to the limited size of the detector the algorithm provides a constantly lower energy measurement for neutrino energies above 50 GeV as most tracks at these energies deposit only a fraction of their total energy within the detector.
 
\newpage

 \textbf{\begin{figure}[H]
 \centering
\includegraphics[width=0.6\linewidth]{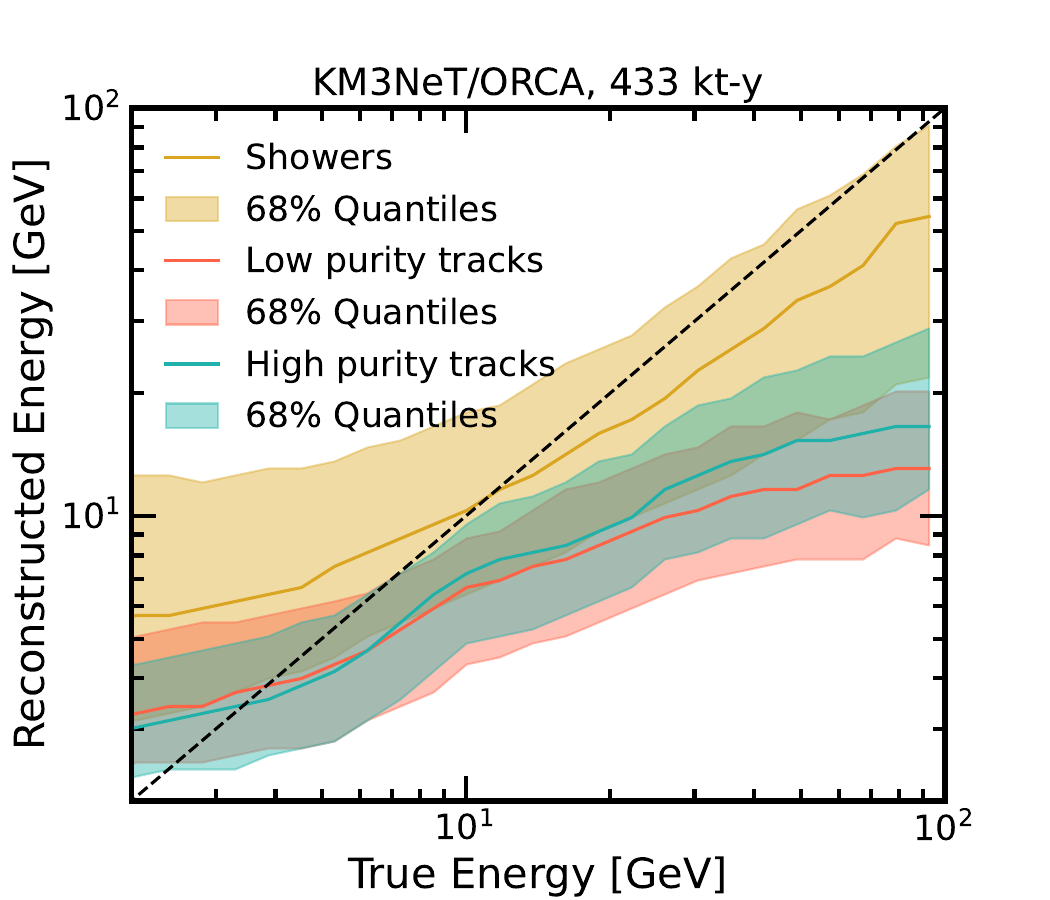}
    \caption{Reconstructed energy as a function of the true energy for High Purity Tracks (blue), Low Purity Tracks (red) and Showers (yellow). The 68\% CL quantiles are drawn and the diagonal line showing a perfect reconstruction.}
    \label{Fig:RecoE}
\end{figure}}
 
In table~\ref{Table:ORCA6Selection_w}, the observed and expected number of selected events per class and interaction channel are shown. Event counts are given after the fit for the atmospheric oscillation parameters is performed and the best fit values are considered (see section~\ref{sec:analysis}). Muon neutrino and antineutrino represent 95\% of the High Purity Tracks sample, 90\% of the Low Purity Tracks sample and 46\% of the Showers sample. The atmospheric muon contamination is respectively 0.4\%, 4\% and 1\%.

\begin{table}[H]
\begin{tabular}{|c|ccc|c|}
\hline
    Channel         & \textbf{High Purity Tracks} & \textbf{Low Purity Tracks} & \textbf{Showers} & \textbf{Total} \\ \hline \hline
$\nu_\mu$ CC                   & 1166.2               & 1187.1               & 670.2              & 3023.5           \\
$\bar{\nu}_\mu$ CC             & 612.4                & 600.8                & 236.0              & 1449.2           \\
$\nu_e$ CC                     & 36.9                 & 62.1                 & 434.5              & 533.5            \\
$\bar{\nu}_e$ CC               & 14.0                 & 22.9                 & 172.5              & 209.4            \\
$\nu_\tau$ CC                  & 14.0                 & 13.0                 & 95.3               & 122.3            \\
$\bar{\nu}_\tau$ CC            & 6.4                  & 5.7                  & 37.3               & 49.4             \\
$\nu$ NC                       & 9.6                 & 16.9                 & 224.3              & 250.8            \\
$\bar{\nu}$ NC                 & 3.0                  & 5.1                  & 66.7               & 74.8             \\
Atm. Muons                     & 7.1                  & 87.5                 & 22.2               & 116.8           \\ \hline
\textbf{Total MC}              & 1869.6               & 2001.1               & 1959.0             & 5829.7           \\ \hline \hline
\textbf{Total Data}            & 1868               & 2002               & 1958             & 5828           \\ \hline
\end{tabular}
\caption{Number of  expected events (MC) and data events passing the selection criteria per class and interaction channel compared to the total number of data events in each class for an  exposure of 433 kton-year. The MC values correspond to the best fit which is detailed in section~\ref{sec:analysis} with nuisance parameters as given in Table~\ref{MINOS_DOscFree}.}
\label{Table:ORCA6Selection_w}
\end{table}

\section{Analysis}
\label{sec:analysis}

The number of expected events is computed using the KM3NeT package \textit{Swim} \cite{Bourret:2018kug}, that combines the atmospheric neutrino flux, neutrino oscillation probabilities and the detector response to calculate the expected number of neutrino events in ORCA. The number of reconstructed events in each class is computed by weighting the simulated events according to the Honda flux \cite{HondaFlux} at the Frejus site without mountain over the detector for solar minima. Oscillation probabilities are computed with the OscProb package \cite{joao_coelho_2023_10104847}. 

The distribution of MC events that pass the selection is built as a 4D-matrix in true energy ($E_{\text{true}}$), true cosine of the zenith angle ($\cos\theta_{\text{true}}$), reconstructed energy ($E_{\text{reco}}$) and reconstructed cosine of the zenith angle ($\cos\theta_{\text{reco}}$) for each interaction channel ($\nu_e$ CC, $\nu_{\mu}$ CC, $\nu_{\tau}$ CC, $\overline{\nu}_{e}$ CC, $\overline{\nu}_{\mu}$ CC,  $\overline{\nu}_{\tau}$ CC, $\nu$ NC, $\overline{\nu}$ NC), repeated for each event class. The binning and ranges for the variables are summarised in table~\ref{Table6-binning}. The binning strategy involves using bins evenly spaced in linear scale for both  the true and reconstructed cosine of the zenith angle. A more complex scheme is used for the reconstructed energy bins:
\begin{itemize}
    \item Initially, equally spaced bins in logarithmic scale between 2 and 100 GeV are built (20 bins).
    \item Subsequently, some bins at the edges are merged to ensure that each bin has more than 2 reconstructed expected events. This requirement ensures the validity of the Barlow-Beeston light method used to estimate the MC statistical uncertainty in the sample, described in section \ref{sec:parameter_es}.
    \item Finally, there is one bin for energies between 100 and 1000 GeV, which is only used in the shower class.
    \item The procedure gives the following bin edges (values in GeV):
    
    [$2.0$, $3.0$, $4.4$, $5.3$, $6.5$, $7.9$, $9.6$, $11.6$, $14.1$, $17.2$, $20.9$, $25.4$, $30.9$, $55.6$, $100.0$, $1000.0$].
\end{itemize}

\begin{table}[H]
    \centering
 \begin{tabular}{|c|c|c|c|c|}
 \hline
     & $E_{\text{true}}$[GeV] & $\cos\theta_{\text{true}}$ & $E_{\text{reco}}$[GeV]  & $\cos\theta_{\text{reco}}$\\ \hline
    Bins & 40 & 80 &15 &10  \\ \hline
    Range &[$1$, $10000$] & [$-$1, 1] & [2, 1000] & [$-$1, 0] \\ \hline
\end{tabular}
    \caption{Number of bins and range in the $\cos\theta$ and energy (true and reconstructed values) for the MC-based response matrix used in this analysis.}
    \label{Table6-binning}
\end{table}

The atmospheric neutrino flux is computed at the centre of each bin. For oscillations, the probabilities are averaged over the bin energy to account for high-frequency oscillations in the bin range. The energy-averaged oscillation probabilities are assumed to be slowly varying with direction and are simply computed at the centre of each bin in $\cos\theta_{\text{true}}$.

A relatively coarse binning was chosen in true energy to improve the computational efficiency of the analysis. A total of 40 bins of true energy evenly distributed in log-scale was found to be sufficient to describe the detector energy response with negligible impact on the predicted event distributions when compared to finer binning options.

\subsection{Parameter estimation}
\label{sec:parameter_es}
The main goal of this analysis is to measure the atmospheric oscillation parameters, i.e. $\theta_{23}$ and $\Delta m^2_{31}$. The best estimate for these parameters is obtained by minimising the following log-likelihood ratio \cite{Cousins2013GeneralizationOC, BAKER1984437}:
\begin{align}
     \lambda(\vec{\theta},\vec{\epsilon}) =  2 \sum_{i}  \Biggl[ (\beta_{i} N^{\text{mod}}_{i}(\vec{\theta} ; \vec{\epsilon})-N^{\text{dat}}_{i})+N^{\text{dat}}_{i} \ln \left( \frac{N^{\text{dat}}_{i}}{\beta_{i}N^{\text{mod}}_{i}(\vec{\theta};\vec{\epsilon})}\right)\Biggr] + \frac{(\beta_{i}-1)^2}{\sigma_{\beta i}^2}  + \nonumber\\+ \sum_k \left(\frac{\epsilon_k-\langle\epsilon_k\rangle}{\sigma_{\epsilon k}}\right)^2, 
    \label{ORCA6_eq_sat}
\end{align}
\noindent where $\vec{\theta}=\{ \theta_{23}, \Delta m^2_{31}\}$. For a given bin $i$,  $N^{\text{mod}}_{i}$ and $N_{i}^{\text{dat}}$ represent the expected number of reconstructed events according to the model and  the number of observed events respectively. The vector $\vec{\epsilon}$ represents the nuisance parameters. Some of them are externally constrained by other experiments. This information enters the log-likelihood as a Gaussian term derived from the PDF of the auxiliary measurement. The mean value, $\langle \epsilon_k\rangle$, and standard deviation $\sigma_{\epsilon k}$ are the parameters used to define these PDFs. The $\beta_i$ coefficients are added to account for uncertainties due to limited MC statistics following the Barlow and Beeston light method \cite{BARLOW1993219, Conway:2011in}. Here, the $\beta_{i}$ are assumed to be normally distributed, $\beta_{i} \sim N(1,\sigma_{\beta i})$ in each bin. Mathematically, the $\beta_{i}$ parameters act as nuisance parameters constrained
by uncorrelated Gaussian priors, modelling bin-by-bin uncorrelated uncertainties. Each coefficient can be minimised analytically, and is given by:
\begin{equation}
    \beta_i =\frac{1}{2}\left[1-N^{\text{mod}}_{i}\sigma_{\beta i}^2+\sqrt{(1-N^{\text{mod}}_{i}\sigma_{\beta i}^2)^2+4N^{\text{dat}}_{i}\sigma^2_{\beta i}}\right].
\end{equation}
\noindent Consequently, applying the light method implies that the only addition to the likelihood evaluation is the knowledge of the uncertainties of the bin contents, $\sigma_{\beta i}$. Ideally, these should be determined by running the entire chain of simulations, reconstructions, and classifications multiple times, with different random seeds for the simulation processes. However, this process to estimate the variance can be computationally expensive. Therefore, the $\sigma_{\beta i}$ values are approximated as the error of the sum of the event weights. This procedure has been validated by comparing to the results of a second method that estimates this uncertainty by bootstrapping the original simulated, using the same MC to draw samples with replacement to recompute the detector response.

\subsection{Nuisance parameters}
\label{sec:nuisance}
Those oscillation parameters for which KM3NeT/ORCA has no sensitivity are fixed to values from external measurements. These are given in table~\ref{Table2-NufitO6}. Nuisance parameters used in this analysis are listed below.

\begin{enumerate}
    \item Normalisation uncertainties act as a scaling factor accounting for selection efficiency mismodelling and cross-section uncertainties:
    \begin{itemize}
        \item The overall normalisation of selected events, $f_{\text{all}}$.
        \item The normalisation of the High Purity Track class, $f_{\text{HPT}}$.
        \item The normalisation of the Shower class, $f_{\text{S}}$.
        \item The normalisation of the NC events, $f_{\text{NC}}$, accounting for the uncertainty in modelling the NC event selection.
        \item The normalisation of the $\tau$ CC events, $f_{\tau \text{CC}}$, accounting for the uncertainty in modelling the $\tau$ CC event selection.
        \item The normalisation of the atmospheric muon background, $f_{\mu}$.
        \item The normalisation of high-energy simulated events, $f_{\text{HE}}$. This normalisation is introduced to take into account the different assumptions on light propagation made by the two light propagation software packages used. A scaling is applied for NC events with true energy above 100~GeV and for CC events with true energy above 500~GeV. The assumed uncertainty is derived from simulations.
    \end{itemize}
    \item Flux shape systematics are applied to the neutrino flux, while the integral of the flux is kept constant. The uncertainties are motivated in refs. \cite{PhysRevD.74.094009, Evans_2017}, with the spectral index uncertainty being conservatively increased to account for a possible contribution from selection efficiency mismodelling.
    \begin{itemize}
        \item The ratio of up-going to horizontally-going neutrinos, applied as $\phi^*(\theta,E)=\phi(\theta,E)\times(1+\delta_{\theta}|\cos\theta|)$, where $\theta$ is the true zenith angle and $E$ is the true neutrino energy, $\phi$ is the nominal flux and $\delta_{\theta}$ is the fitted parameter. 
        \item   The spectral index of the neutrino flux, applied as $\phi^*(\theta,E)=\phi(\theta,E)\times E^{\delta_{\gamma}}$, where $\delta_{\gamma}$  is the fitted parameter.
        \item The ratio of muon neutrinos to muon antineutrinos, the ratio of electron neutrinos to electron antineutrinos, and the ratio of muon neutrinos and electron neutrinos. They are applied as:
        \begin{align}   
            \delta_e = (1+s_{e\mu}) \cdot (1+s_{e\overline{e}})\\
            \delta_{\overline{e}} = (1+s_{e\mu} )\cdot (1-\frac{I_{e}}{I_{\overline{e}}}s_{e\overline{e}})\\
            \delta_{\mu}=(1-\frac{I_{e}+I_{\overline{e}}}{I_{\mu}+I_{\overline{\mu}}}s_{e\mu})\cdot (1+s_{\mu\overline{\mu}})\\
            \delta_{\overline{\mu}}=(1-\frac{I_{e}+I_{\overline{e}}}{I_{\mu}+I_{\overline{\mu}}}s_{e\mu})\cdot (1-\frac{I_{\mu}}{I_{\overline{\mu}}}s_{\mu\overline{\mu}}),
    \end{align}
    where $\delta_e$, $\delta_{\mu}$, $\delta_{\bar{e}}$, $\delta_{\bar{\mu}}$ are the individual normalisations of each component of the flux, $I_e$, $I_{\mu}$, $I_{\bar{e}}$ and $I_{\bar{\mu}}$ are the nominal integrals of each flux component and $s_{\mu\overline{\mu}}$, $s_{e\overline{e}}$ and $s_{e\mu}$ are the fitted ratio parameters.  
    \end{itemize}

   \item The absolute energy scale of the detector, $E_{s}$. This parameter accounts for the uncertainty on water optical properties, knowledge of PMT efficiencies, and knowledge on the hadronic shower modelling, amounting to a global 9\% uncertainty. This value is applied as a shift in the true energy of the detector response \cite{chau:tel-03999509, Adri_n_Mart_nez_2016}.
\end{enumerate}

\begin{table}[H]
\centering
    \begin{tabular}{|l|c|}
      \hline
      \textbf{Parameter} & \textbf{NO} \\
      \hline
      $\theta_{12}$ & $33.44^{\circ}$ \\
      \hline
      $\theta_{13}$  & $8.57^{\circ}$  \\
      \hline
      $\Delta m_{21}^2$ & $7.42 \times 10^{-5}~\mathrm{eV^2} $\\ 
      \hline
      $\delta_{\text{CP}}$ & $197^{\circ}$ \\
      \hline
     
    \end{tabular}
    \caption{List of the three-flavour neutrino oscillation parameters from NuFit 5.0 \cite{Esteban:2020cvm} for NO including Super-Kamiokande (SK) data that are fixed in the analysis.}
    \label{Table2-NufitO6}
    \end{table}

The best fit is obtained by minimising the negative log-likelihood ratio (Eq. \ref{ORCA6_eq_sat}) using 8 starting points indicated in table~\ref{Table3-StartPoints}.
\clearpage
\begin{table}[H]
\centering
    \begin{tabular}{|c|c|c|}
      \hline
      $\theta_{23}$ [$^{\circ}$] &$ \Delta m^2_{31}$ [$\times10^{-3}~\mathrm{eV^2}$] & $E_s$\\
      \hline
      $40$ & 2.517 & 0.95 \\ \hline
      $40$ & 2.517 & 1.05 \\ \hline
      $40$ &-2.428 & 0.95 \\ \hline
      $40$ &-2.428 & 1.05  \\ \hline
      $50$ & 2.517 & 0.95  \\ \hline
      $50$ & 2.517 & 1.05 \\ \hline
      $50$ &-2.428 & 0.95 \\ \hline
      $50$ &-2.428 & 1.05 \\ \hline
     
    \end{tabular}
    \caption{Starting values of the parameters for each fit. The parameter space is restricted to the corresponding $\theta_{23}$ octant, mass ordering and energy scale below/above 1.0 in each fit. }
    \label{Table3-StartPoints}
    \end{table}

\subsection{Pseudo-experiments}

Pseudo-experiments (PEs) are generated to extract the expected distributions of the test statistics. In this analysis, the approach to generating pseudo-experiments is the following:
\begin{itemize}
\item Nuisance parameters are set to their nominal values.
\item Oscillation parameters are set to the NuFit 5.0 values with Super-Kamiokande data as given in table~\ref{Table2-NufitO6}. NO is assumed unless mentioned otherwise.
\item The mean values $\langle\epsilon\rangle$ of the nuisance parameters that are constrained by external data are sampled according to their distributions. This approach acknowledges the frequentist nature of auxiliary measurements, where the true values of these parameters are fixed. In this sense, the auxiliary data is treated in the same way as the analysed data \cite{Cranmer:2014lly}. 
\item The bin contents are sampled according to Poisson statistics.
\end{itemize}

\section{Results}
\label{sec:results}

\begin{figure}[H]
    \centering    \includegraphics[width=0.7\linewidth]{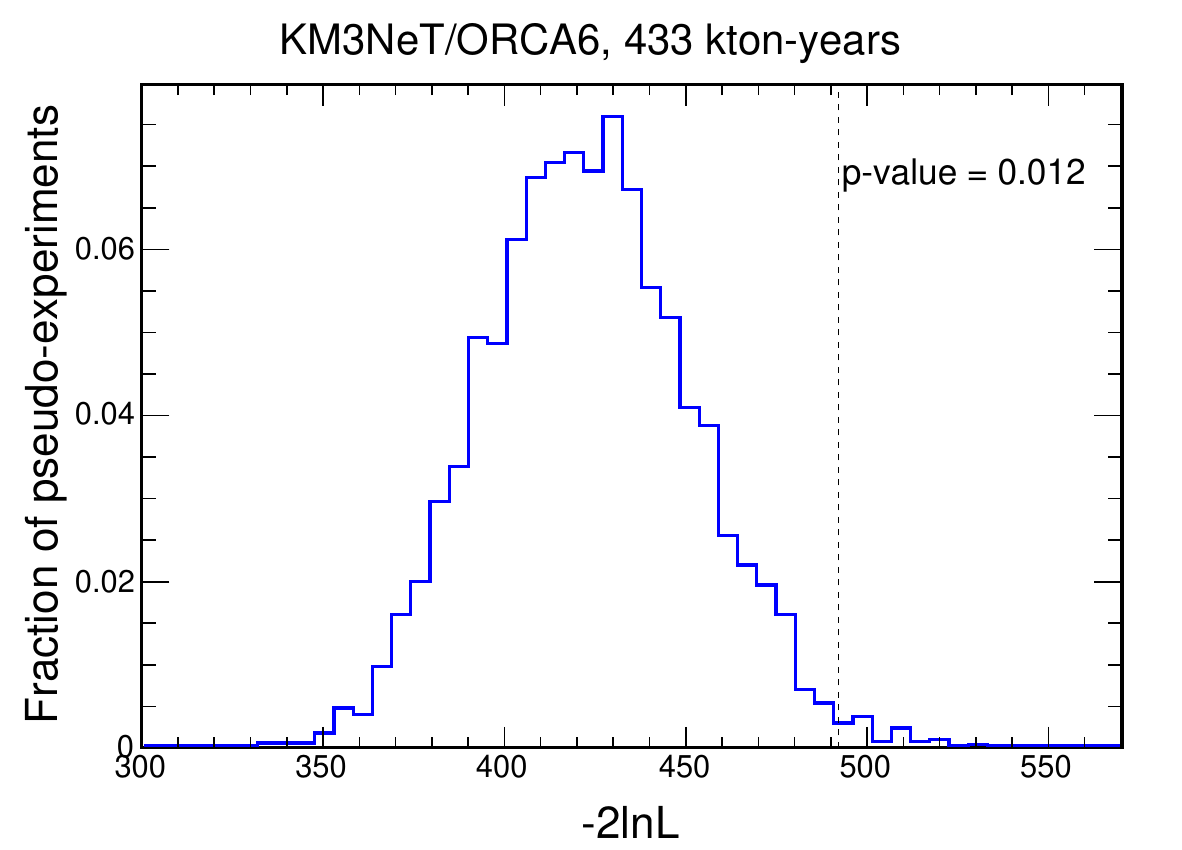}
    \caption{Distribution of $\lambda$ values from 5000 PEs used to carry out the goodness-of-fit test. The vertical line indicates the value of the data best-fit log-likelihood.}
    \label{Fig:C6-GoF}
\end{figure}

After performing the full minimisation of equation \eqref{ORCA6_eq_sat}  the observed log-likelihood ratio $\lambda$ is found to be 492.1. To assess the goodness-of-fit, a set of 5000 pseudo-experiments are generated assuming NuFit 5.0 values for the parameters of interest and the nuisance parameters at their nominal values. The log-likelihood ratio $\lambda_{\text{GoF}} = -2\ln L$ is computed for each of them and the corresponding distribution is shown in figure~\ref{Fig:C6-GoF}. The probability of obtaining a $\lambda_{\text{GoF}}$ value equal or larger than the observed $\lambda_{\text{GoF}}$ is found to be  $(1.20~\pm~0.15)\%$. The uncertainty is derived by bootstrapping the 5000 PEs, i.e. sampling the pseudo-experiments with replacement to recompute the p-value.

The best-fit values  are:
\begin{align}
&\sin^2\theta_{23}= 0.51^{+0.04}_{-0.05} \\
     &\Delta m^2_{31} =
    \begin{cases}
      2.18^{+0.25}_{-0.35}\times 10^{-3}~\mathrm{eV^2}, \quad \text{for NO}. \\
      [-2.25,-1.76]\times 10^{-3}~\mathrm{eV^2},\quad \text{for IO}
    \end{cases} \\
&2\ln(L_{\text{NO}}/L_{\text{IO}}) = 0.31
\end{align}

The errors were computed via the Feldman-Cousins method \cite{Feldman_1998} for 68\% CL and incorporate both the statistical and systematic uncertainties. 

\begin{figure}[H]
 \centering
      \includegraphics[width=0.49\linewidth]{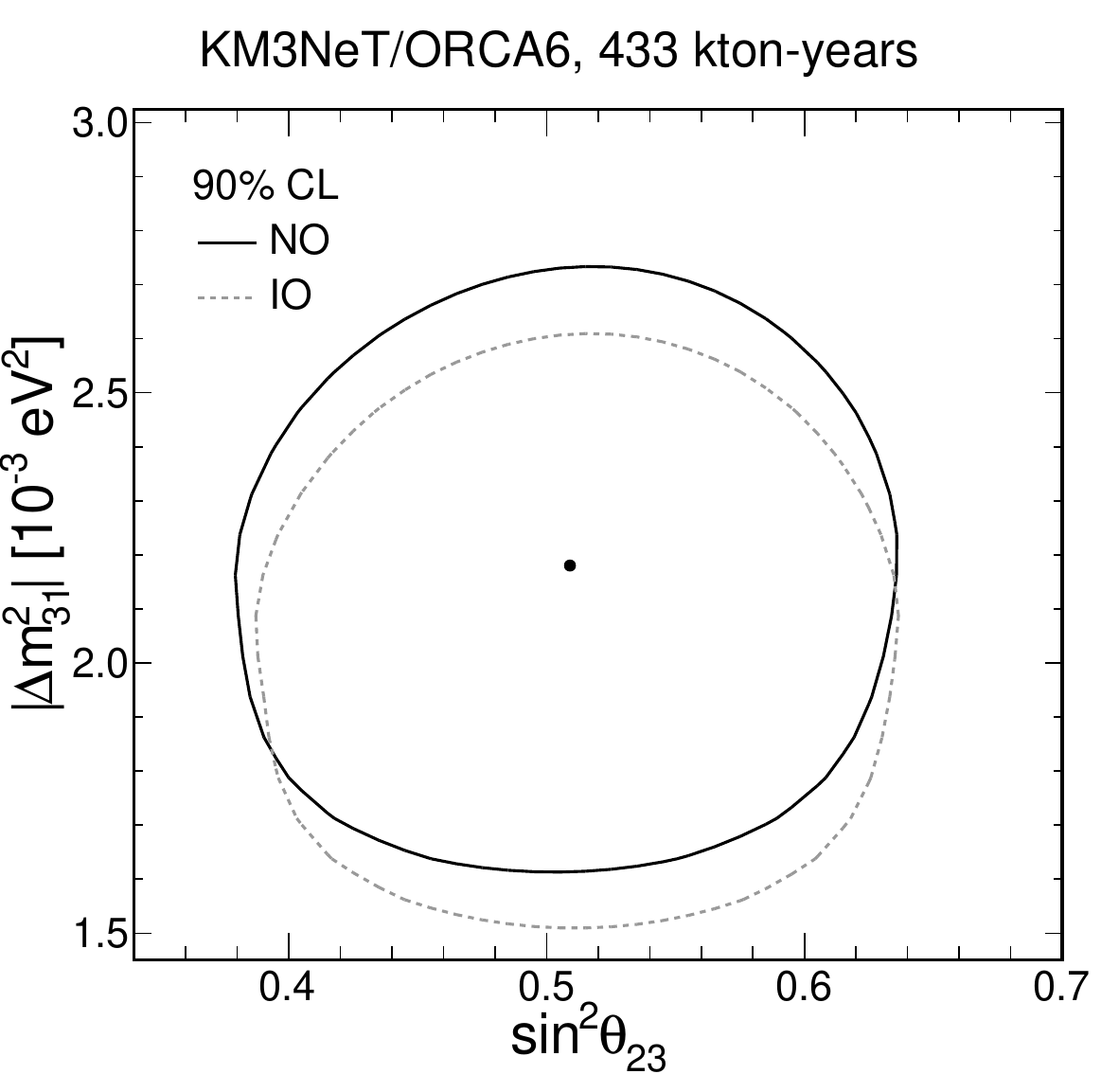}\includegraphics[width=0.49\linewidth]{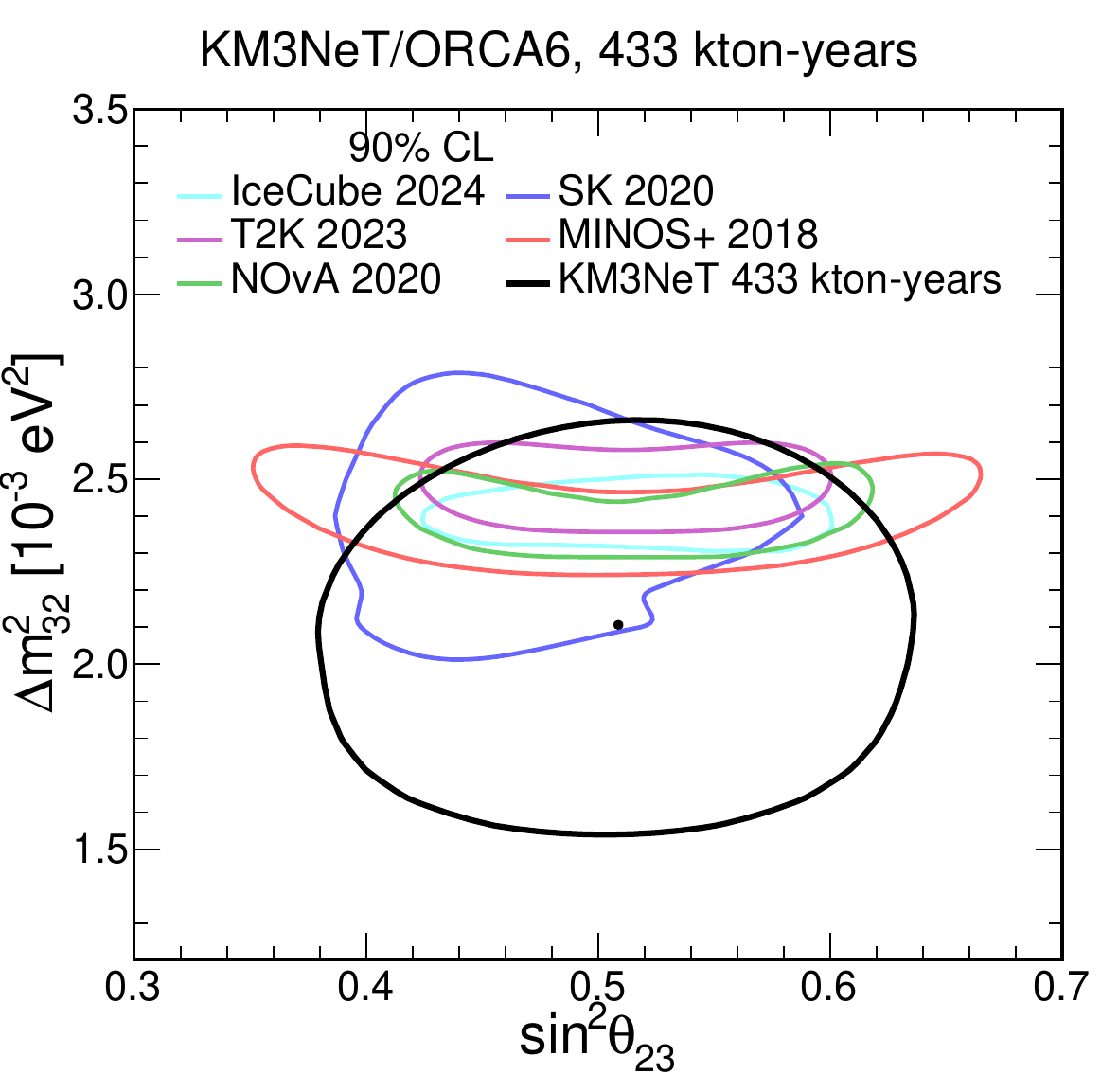}
 \caption{\textbf{Left}: Allowed regions at at 90\% CL obtained from ORCA6 data for the oscillation parameters $\sin^2 \theta_{23}$ and $\Delta m^2_{31}$. The best fit is indicated with a dot. The NO (IO) is depicted in solid black (dashed black). 
 \textbf{Right}: Allowed region at 90\% CL for ORCA6 (black solid line) assuming NO for the oscillation parameters $\sin^2 \theta_{23}$ and $\Delta m^2_{32}$ compared with other experiments:  IceCube \cite{icecubecollaboration2024measurement}, T2K \cite{abe2023updated}, Super-Kamiokande (SK), \cite{yasuhiro_nakajima_2020_4134680}, MINOS+ \cite{aurisano_2018_1286760}, and NO$\nu$A \cite{alex_himmel_2020_4142045}.} 
 \label{c6-contour}
\end{figure}

The allowed region for both oscillation parameters ($\theta_{23}$ and $\Delta m^2_{31}$) at 90\% CL is shown in figure~\ref{c6-contour} (left). The contours are derived according to Wilks' theorem \cite{s__s__wilks_1938} due to computing constraints. In figure~\ref{c6-contour} (right), the NO solution is compared to other experiments. Note that a conversion has been done in this figure as $\Delta m^2_{32} = \Delta m^2_{31}-\Delta m^2_{21}$, with a value for $\Delta m^2_{21}=7.42\times 10^{-5}~\mathrm{eV^2}$.
\clearpage

\begin{figure}[H]
 \centering
     \includegraphics[width=0.49\linewidth]{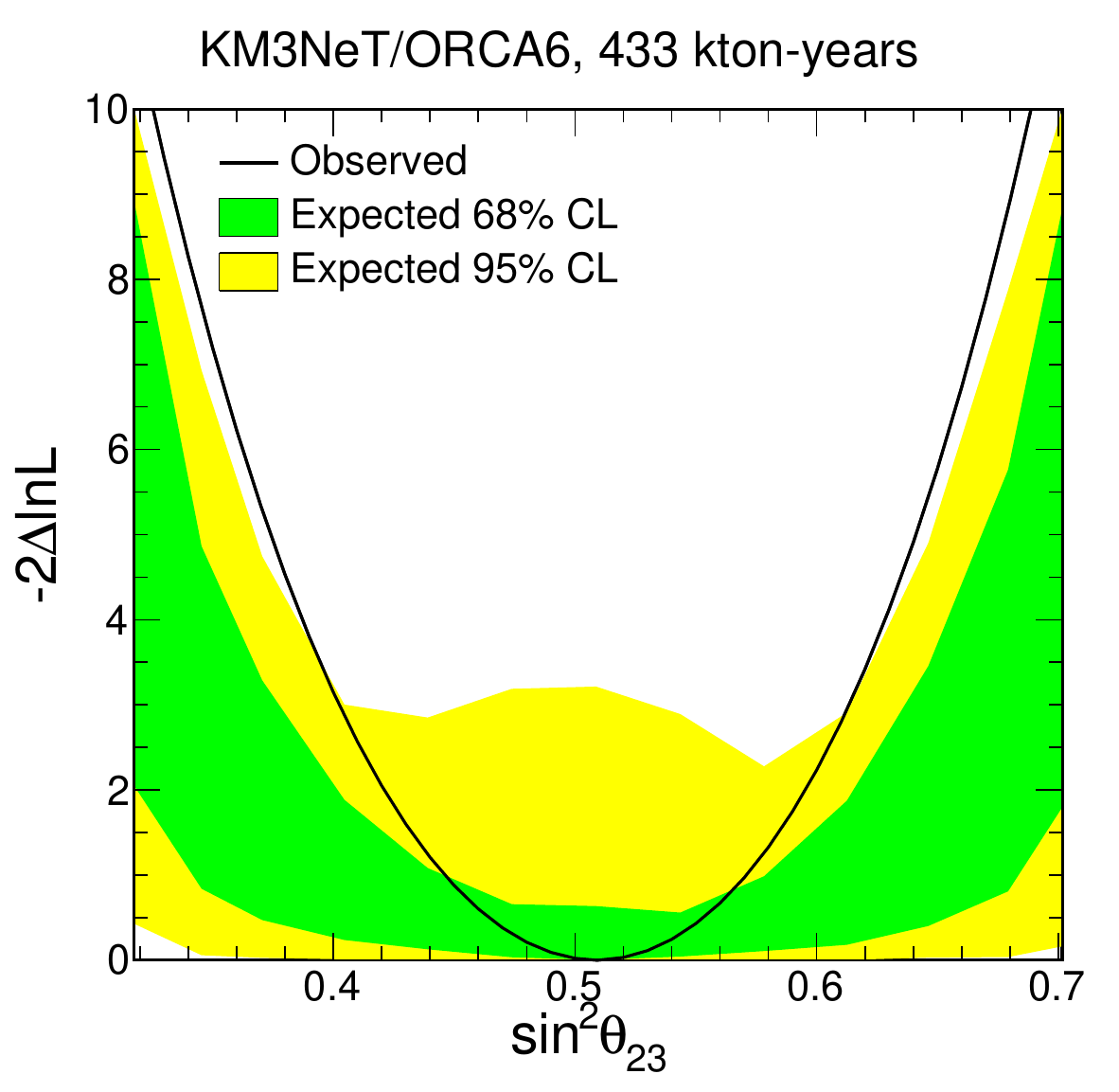}\\
      \includegraphics[width=0.49\linewidth]{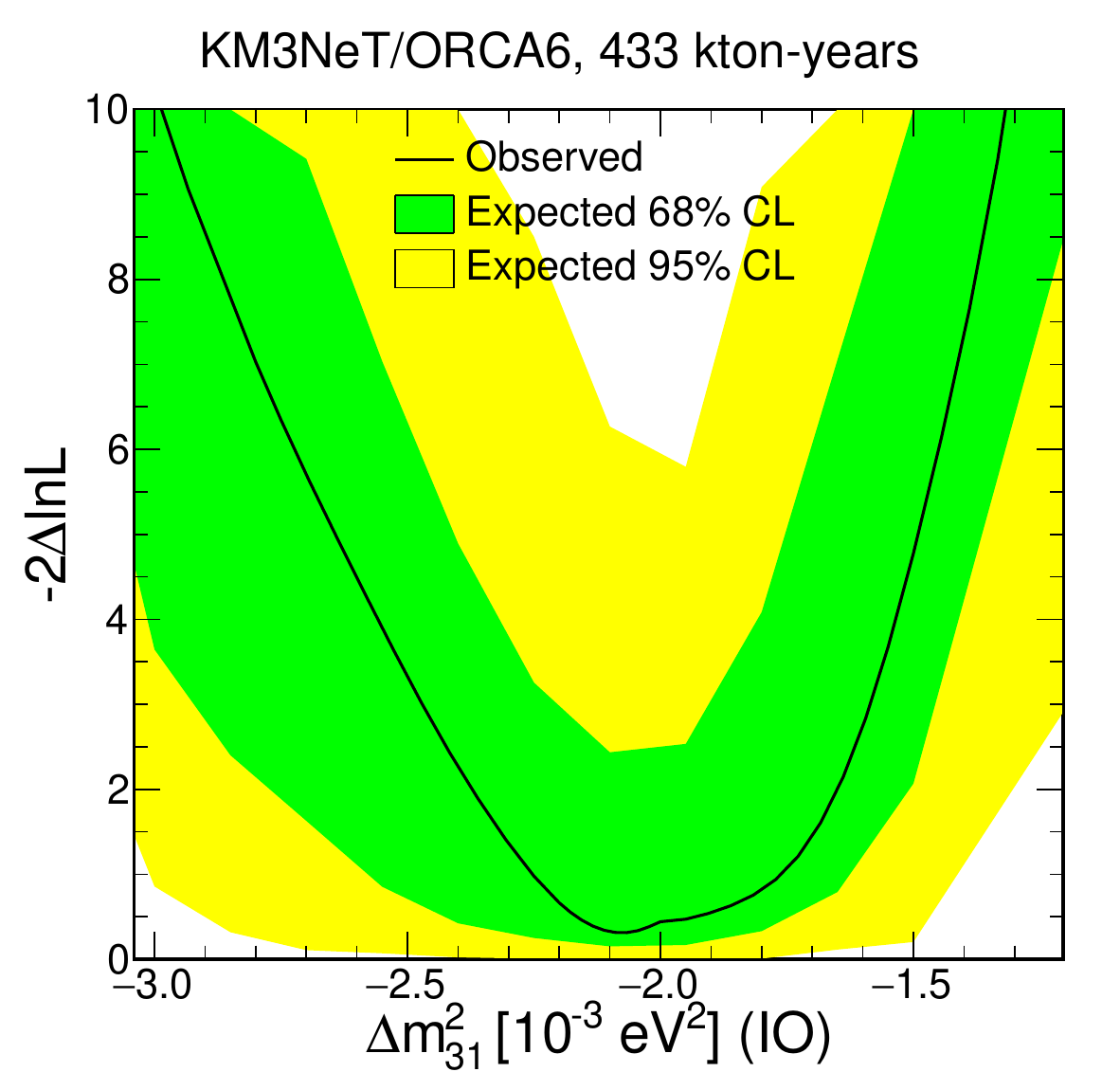}\includegraphics[width=0.49\linewidth]{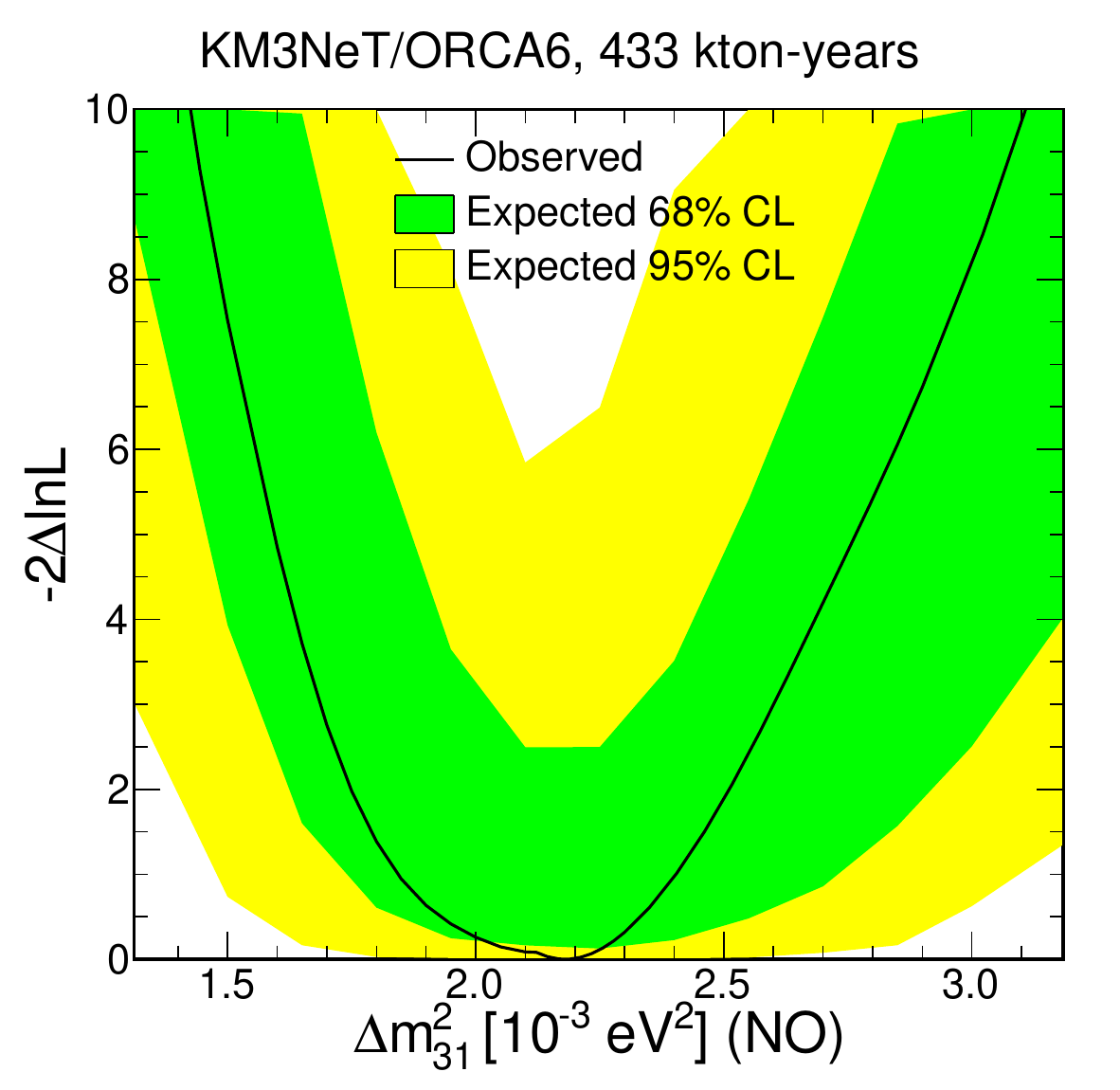}
 \caption{Observed profiled log-likelihood scan (black) and the distribution of 2000 pseudo-experiments (yellow and green bands) produced at the best-fit point for the oscillation parameters, $\sin^2\theta_{23}$ (top) and $\Delta m^2_{31}$ (bottom) for IO (left) and NO (right).}
 \label{Fig:Profiles}
\end{figure}

In figure~\ref{Fig:Profiles}, the profiled log-likelihood ratio scans, $-2\Delta\ln L = -2[\ln L(\vec{\theta}, \hat{\hat{\vec{\epsilon}}})-\ln L({\hat{\vec{\theta}},\hat{\vec{\epsilon}}})]$, are displayed  for $\vec{\theta}=\theta_{23}$ and $\vec{\theta}=\Delta m^2_{31}$ together with the 68\% and 95\% CL bands. The double hat indicates that the nuisance parameters are profiled, i.e. the values of $\epsilon$ that minimise the negative log-likelihood for the specified $\vec{\theta}$. The bands are computed by generating PEs with the true value of the parameters of interest at the best-fit point and the nuisance parameters at the nominal values. The observed $\sin^2\theta_{23}$ curve lies outside the $95\%$ CL band. This happens because the best-fit value ($\sin^2\theta_{23}=0.51$) implies maximal disappearance, and the data contains fluctuations that favour values beyond maximal. The measurement of $\Delta m^2_{31}$ is within the expected 68\% CL bands.
\clearpage

\begin{figure}[H]
 \centering
     \includegraphics[width=0.49\linewidth]{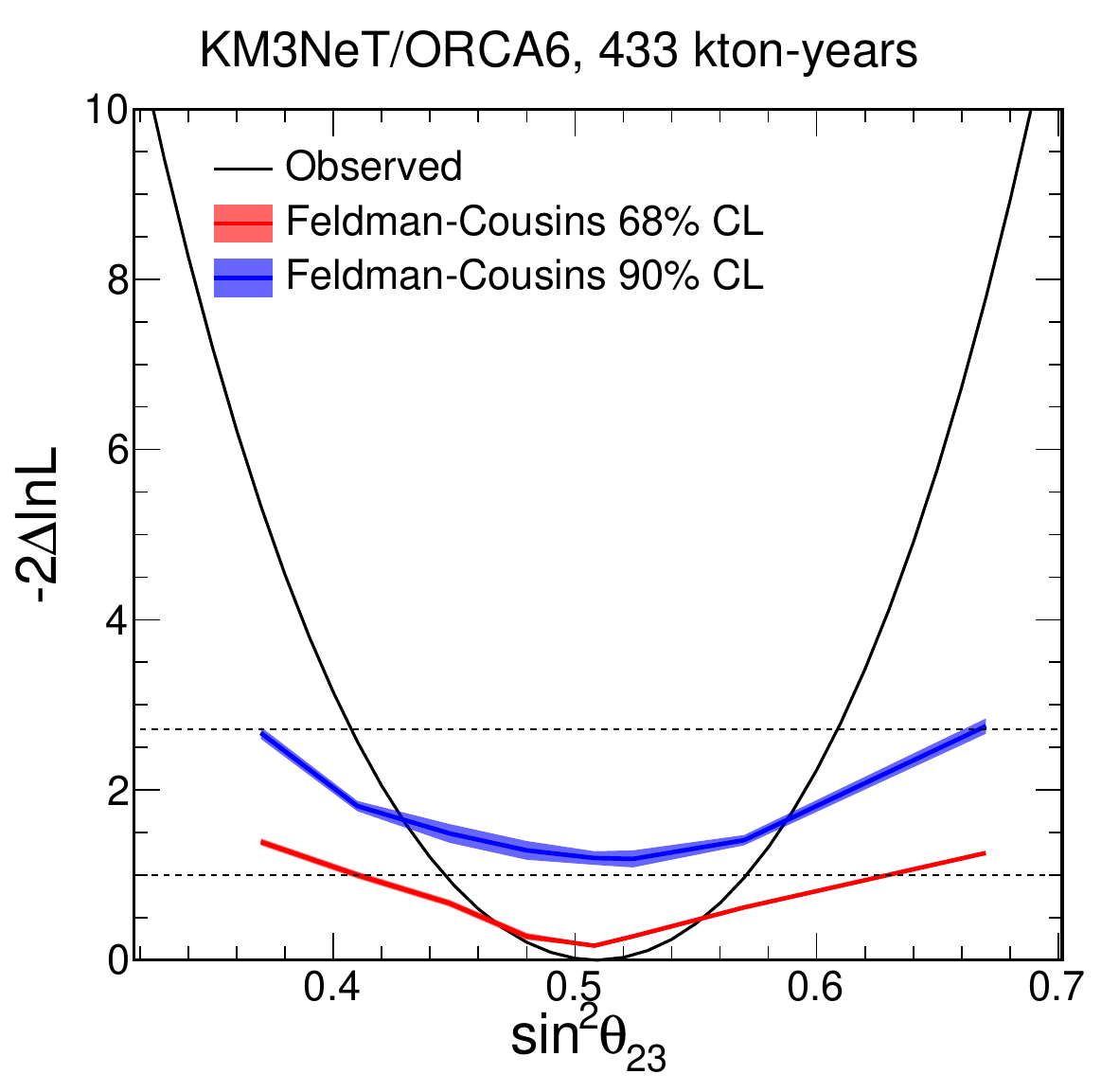} \\
      \includegraphics[width=0.49\linewidth]{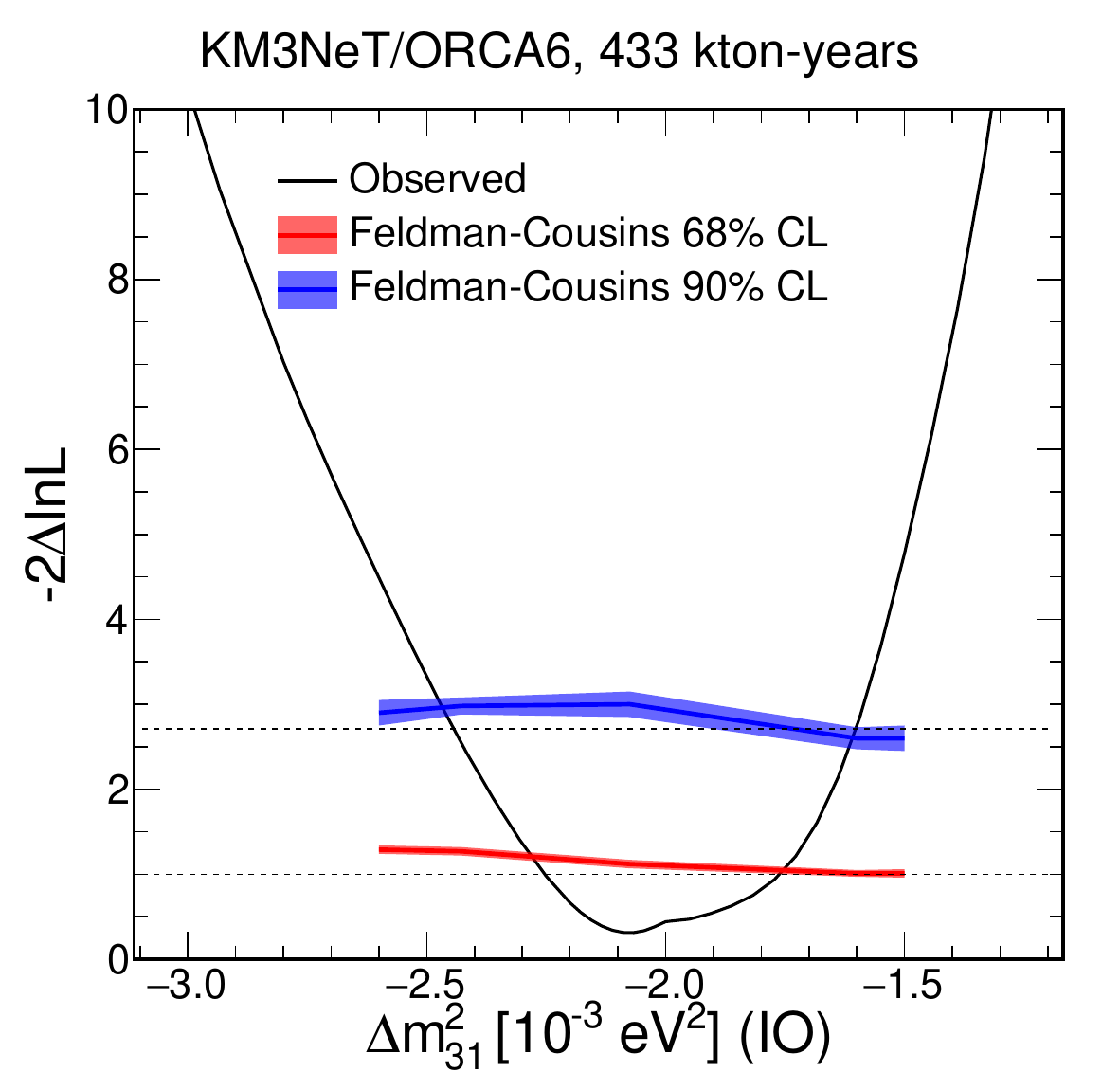}\includegraphics[width=0.49\linewidth]{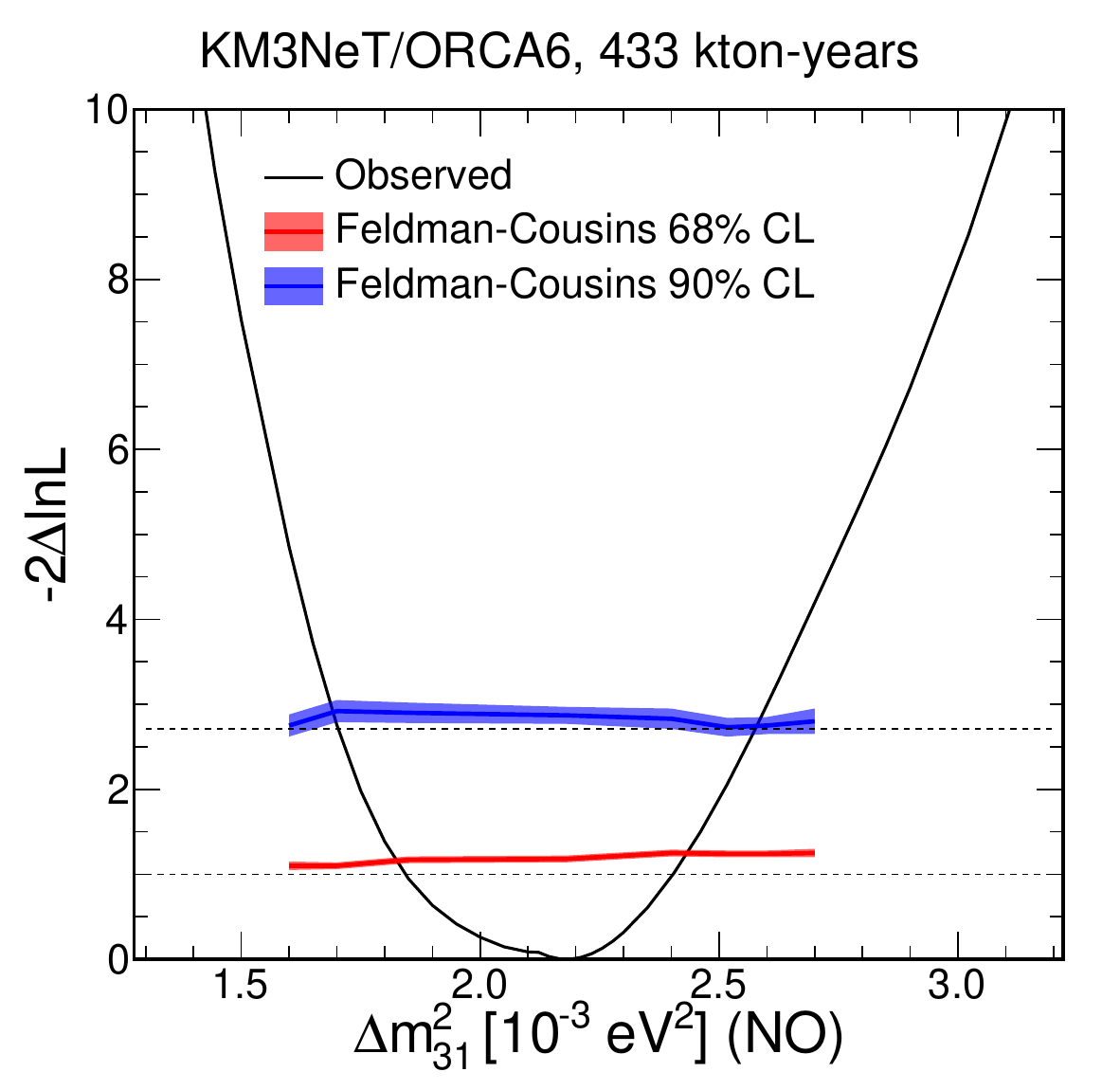}
 \caption{Profiled log-likelihood scan of the oscillation parameters, $\sin^2\theta_{23}$ (top) and $\Delta m^2_{31}$ (bottom) for IO (left) and NO (right). The black curve represents the observed result. Horizontal dashed lines represent the 68\% and 90\% CL thresholds according to Wilks' theorem, while the red and blue bands represent respectively the 68\% and 90\% CL Feldman-Cousins critical values. The uncertainty bands are the standard deviation on the critical values derived by sampling the pseudo-experiments with replacement.}
 \label{C6-StdFC}
\end{figure}

 In order to compute Feldman-Cousins corrections to the parameter uncertainties, a set of 2000 PEs is generated for several testing points. The results compared to Wilks' theorem can be seen in figure~\ref{C6-StdFC}. While the limits on the mass splitting are largely unaffected, the impact on the mixing angle is significant. This effect on the mixing angle is expected, as fluctuations in the data favour values beyond maximal, and the validity conditions of Wilks' theorem are not met due to the presence of boundaries in the parameter space \cite{article}.
\clearpage

\subsection{Neutrino mass ordering significance}

\begin{figure}[H]
    \centering
    \includegraphics[width=0.7\linewidth]{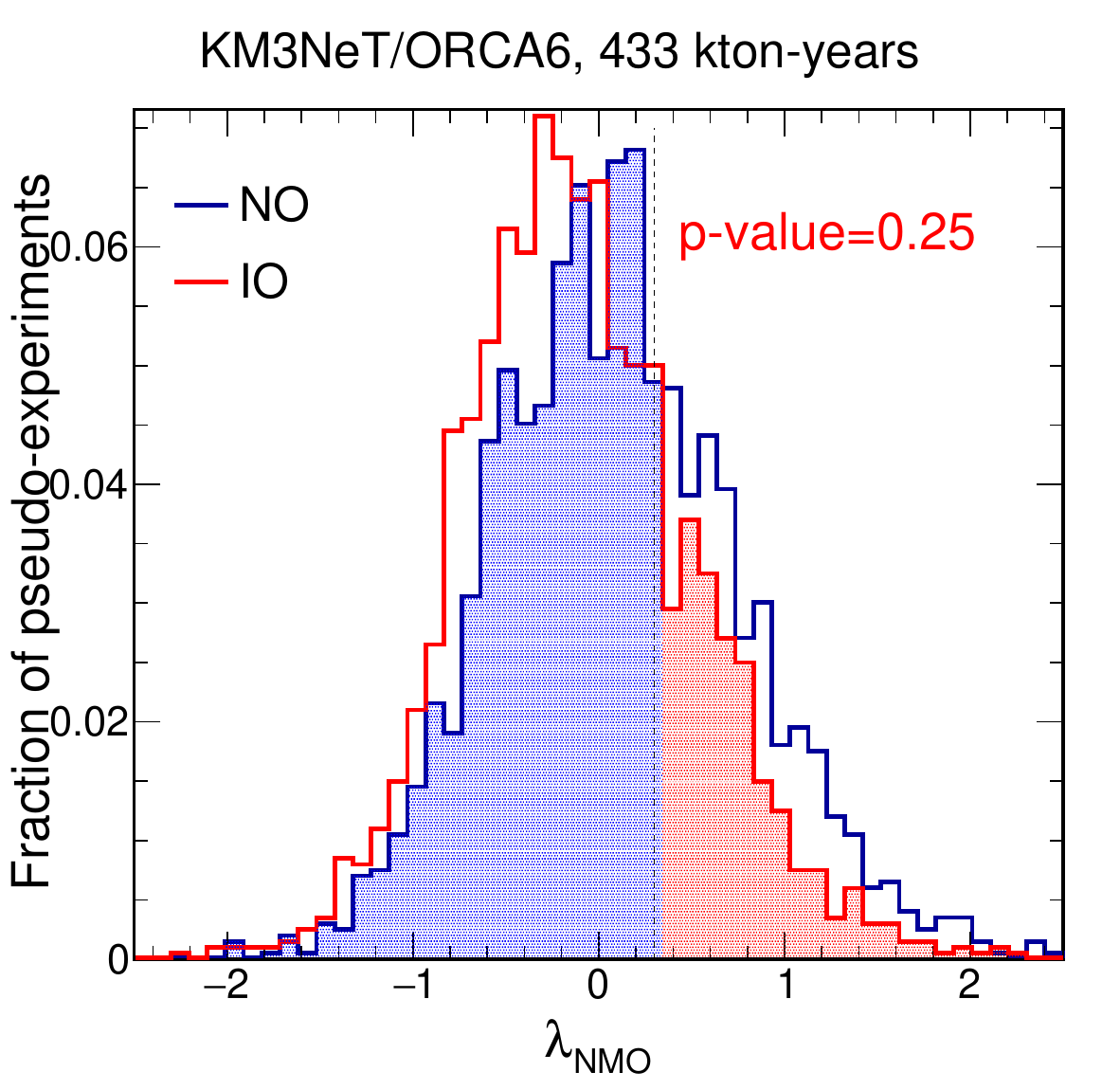}
    \caption{Distribution of the test statistic $\Delta \lambda$, obtained from a set of 2000 pseudo-experiments generated assuming as true the NO (blue) and IO (red) NuFit v5.0 values of $\Delta m^2_{31}$. The vertical line indicates the observed test statistic value. The p-value for IO hypothesis is also depicted.}  \label{Fig:NMODIST}
\end{figure}

The test for neutrino mass ordering is performed using as a test statistic the log-likelihood ratio:
\begin{equation}
    \lambda_{\text{NMO}} = \lambda_{\text{IO}} - \lambda_{\text{NO}}
\end{equation}
A set of 2000 pseudo-experiments, is generated using as true parameters the NuFit v5.0 best-fit values of $\Delta m^2_{31}$ for both mass ordering hypotheses. The test statistic distributions are shown in figure~\ref{Fig:NMODIST}. The value of the observed log-likelihood ratio $\lambda_{\text{NMO}}=0.31$ is indicated with a vertical line. The probability of obtaining a value of $\lambda_{\text{NMO}}$ equal to or larger than the observed $\lambda_{\text{NMO}}$ for the IO hypothesis corresponds to 25\%, while the probability of obtaining a value of $\lambda_{\text{NMO}}$ equal to or smaller than the observed $\lambda_{\text{NMO}}$ for the NO hypothesis is 65\%. Therefore, both hypotheses are compatible with data.

\subsection{Event distributions}
\begin{figure}[H]
 \centering
     \includegraphics[width=0.49\linewidth]{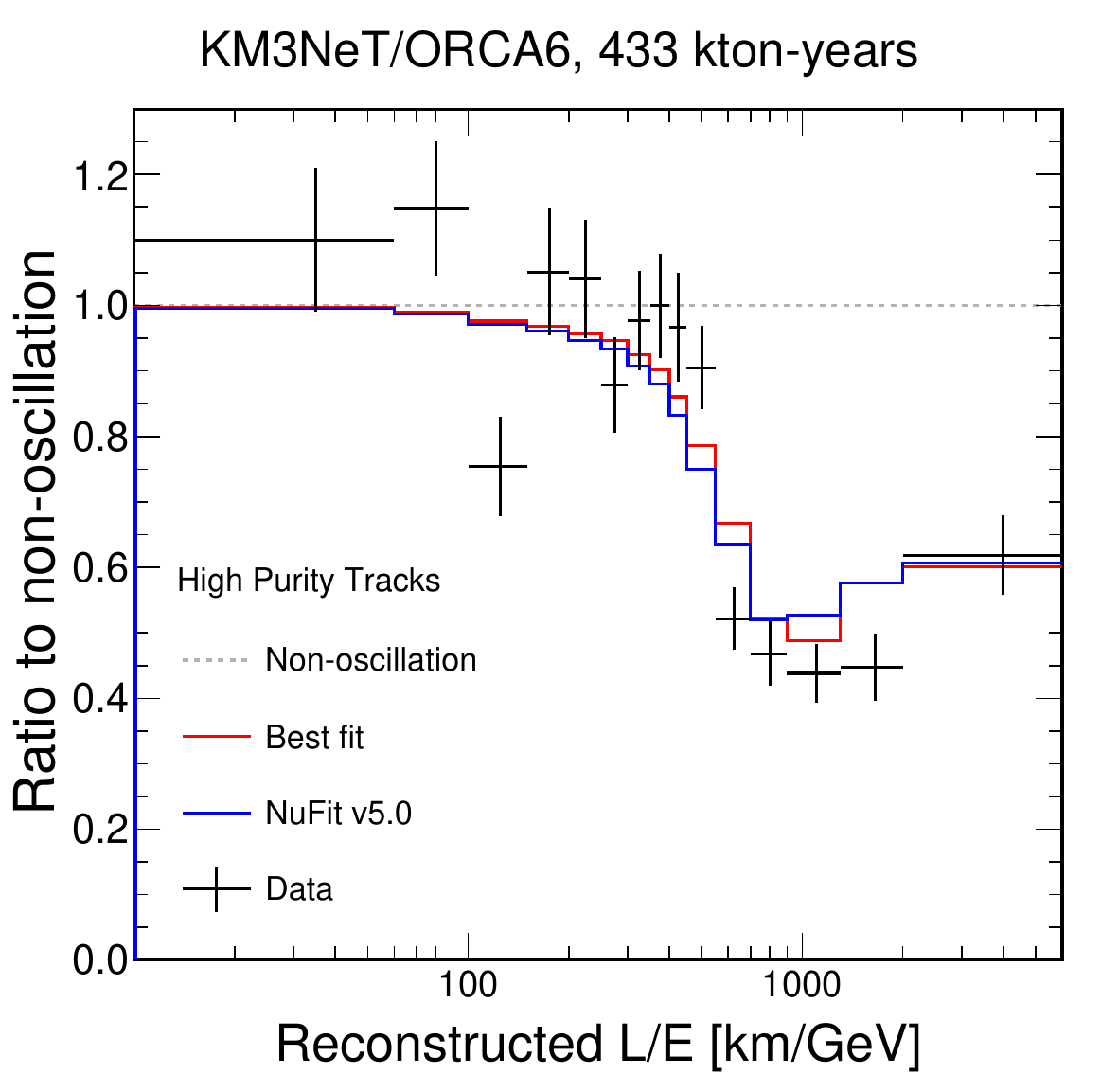}
     \includegraphics[width=0.49\linewidth]{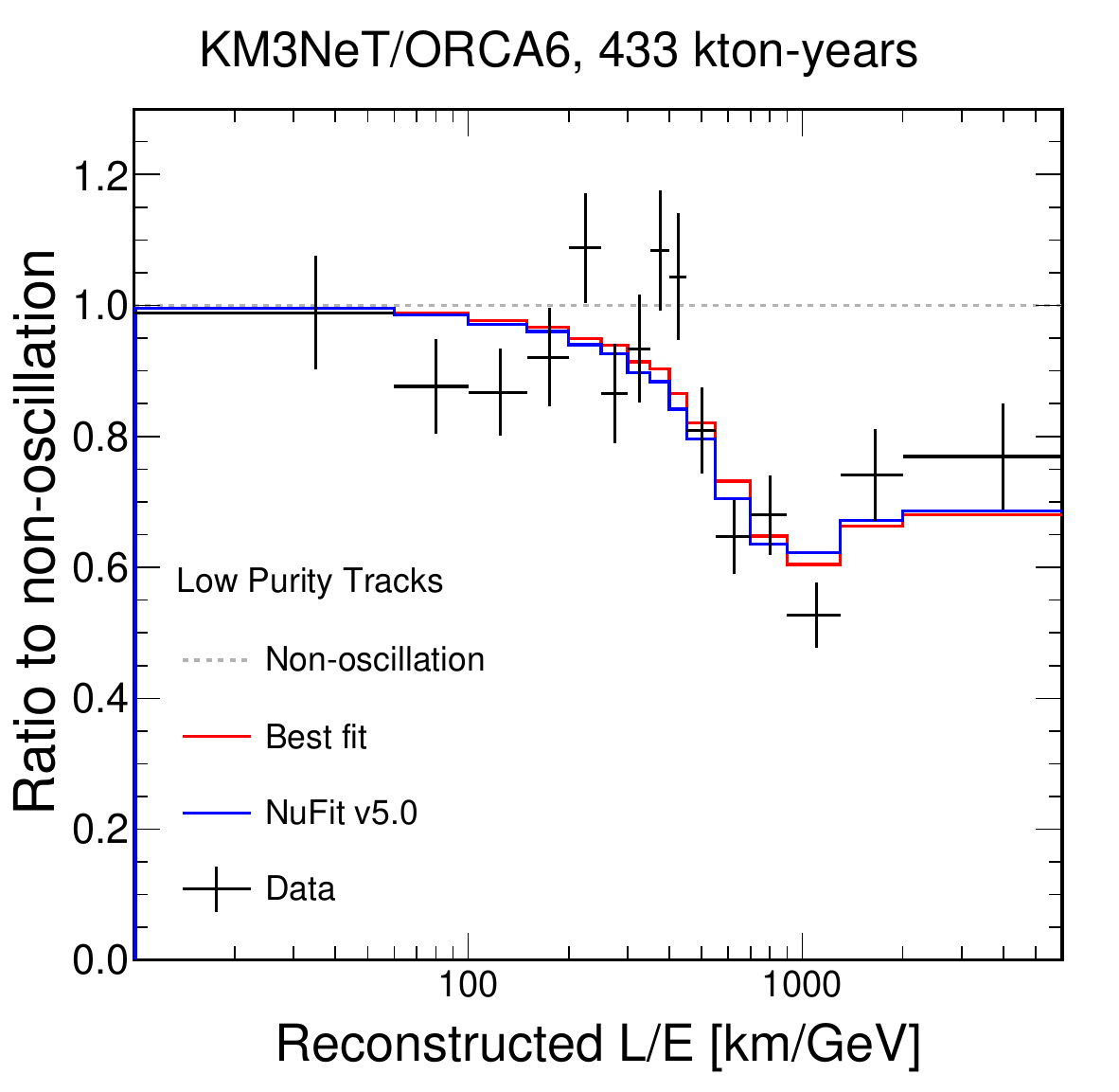} \\
     \includegraphics[width=0.49\linewidth]{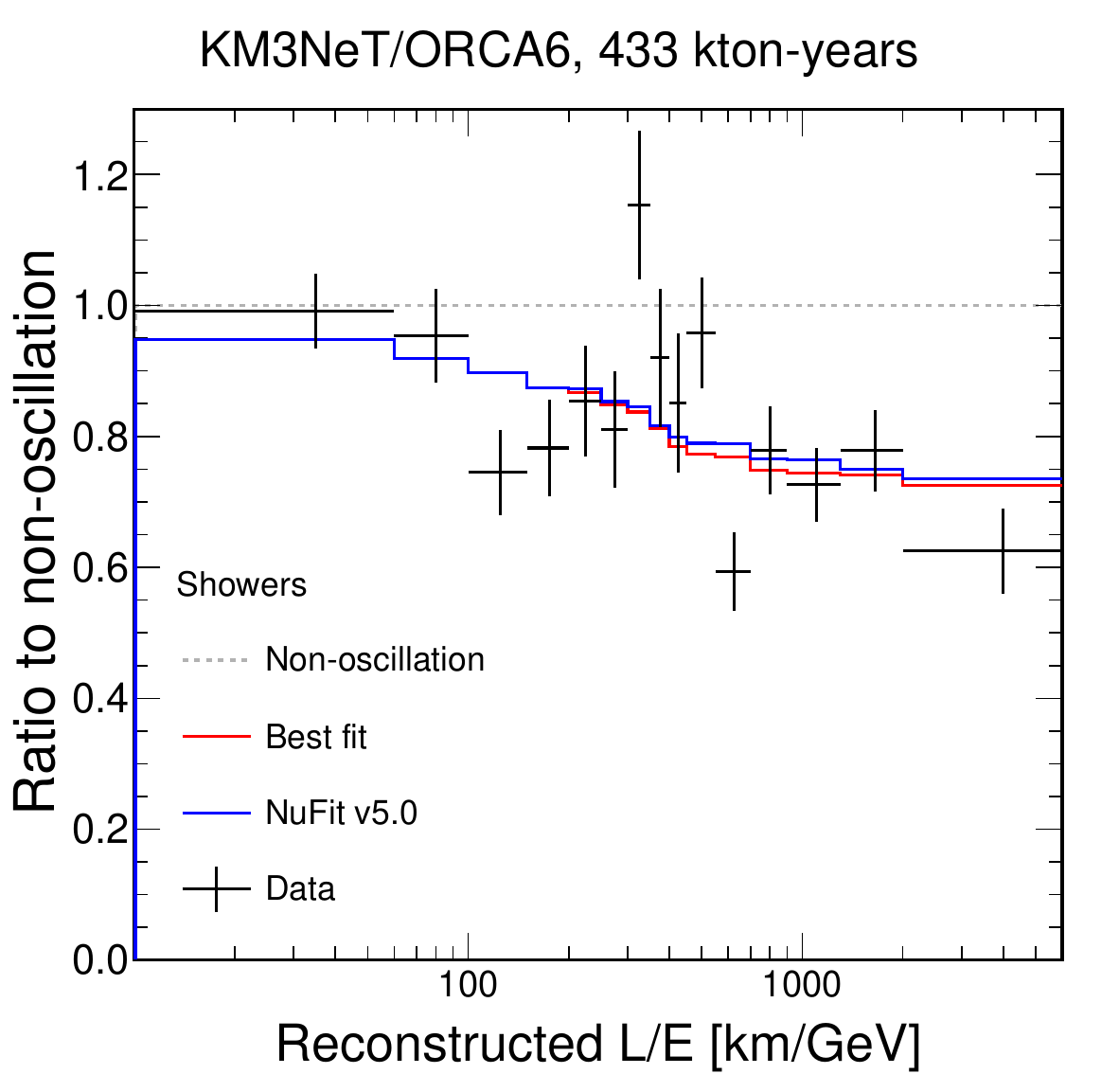}
 \caption{Ratio to non-oscillation hypothesis as a function of the reconstructed  propagation length over energy for High Purity Tracks (top-left), Low Purity Tracks (top-right) and Showers (bottom). The data points are shown with error bars in black, the best fit is shown in solid red and the oscillation hypothesis with NuFit v5.0 values is shown in solid blue. The non-oscillation hypothesis is indicated as a dashed horizontal line.}
 \label{c6-Distr}
\end{figure}

\begin{figure}[]
 \centering
     \includegraphics[width=0.48\linewidth]{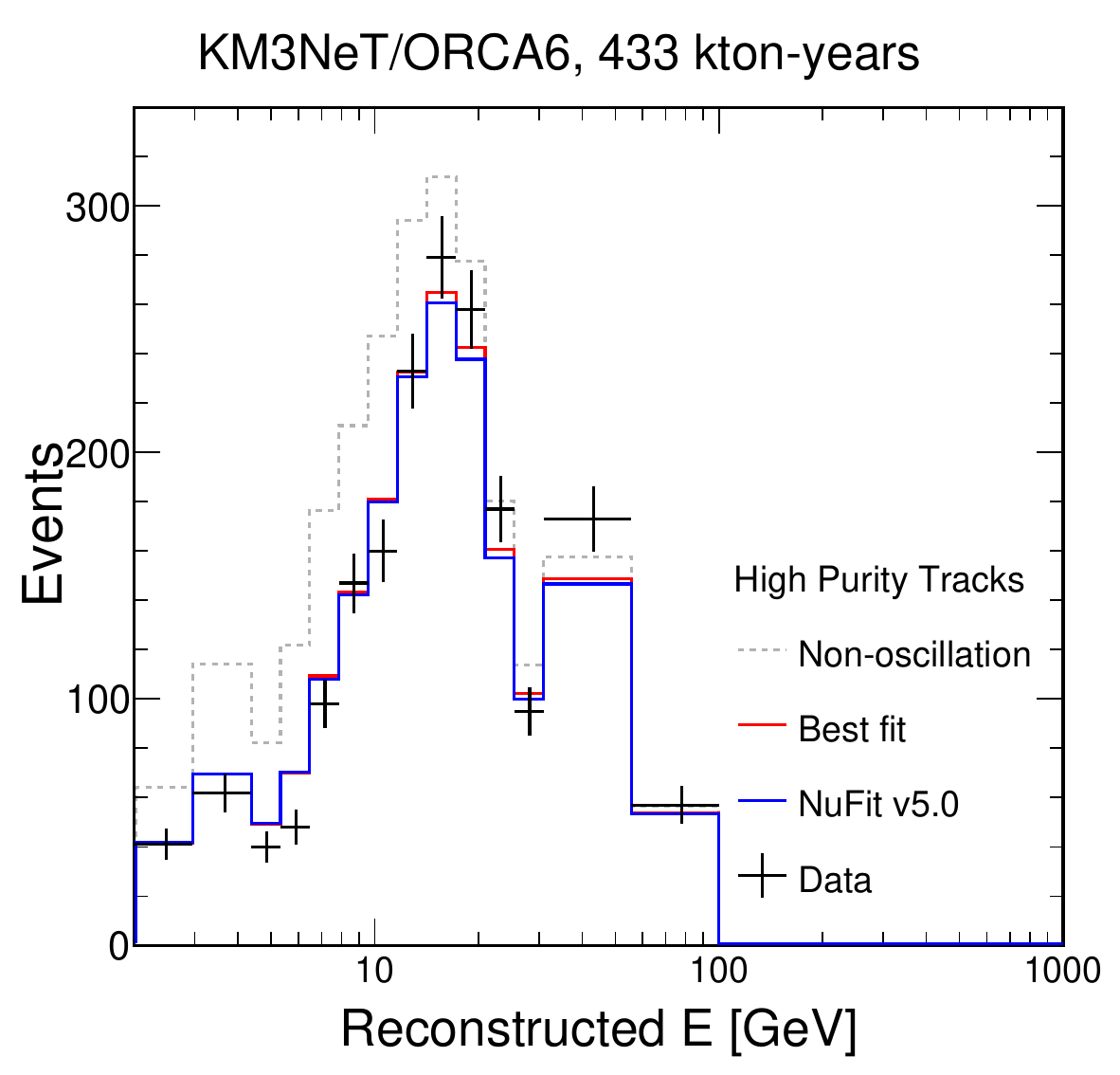}\includegraphics[width=0.48\linewidth]{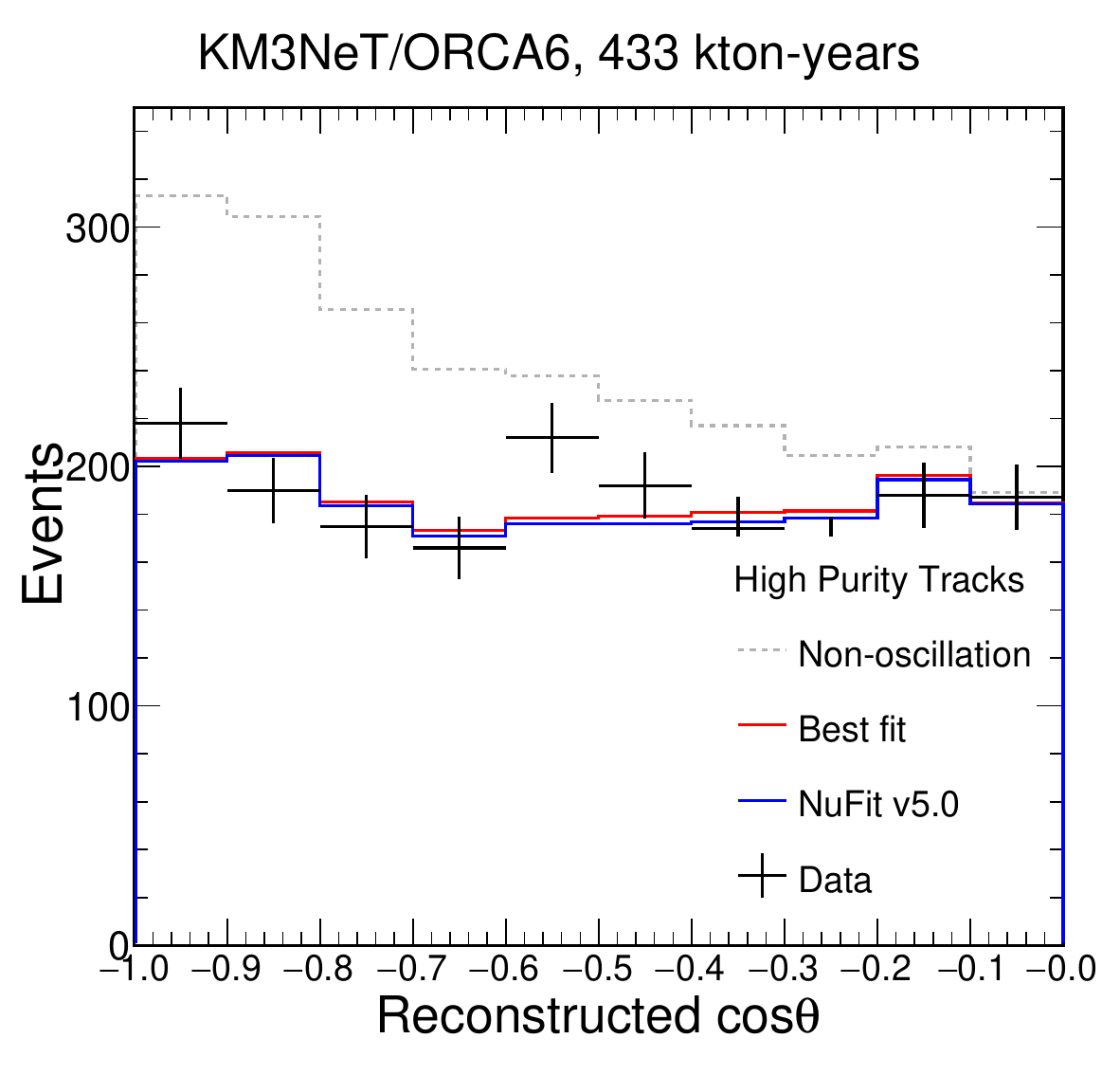} \\
     \includegraphics[width=0.48\linewidth]{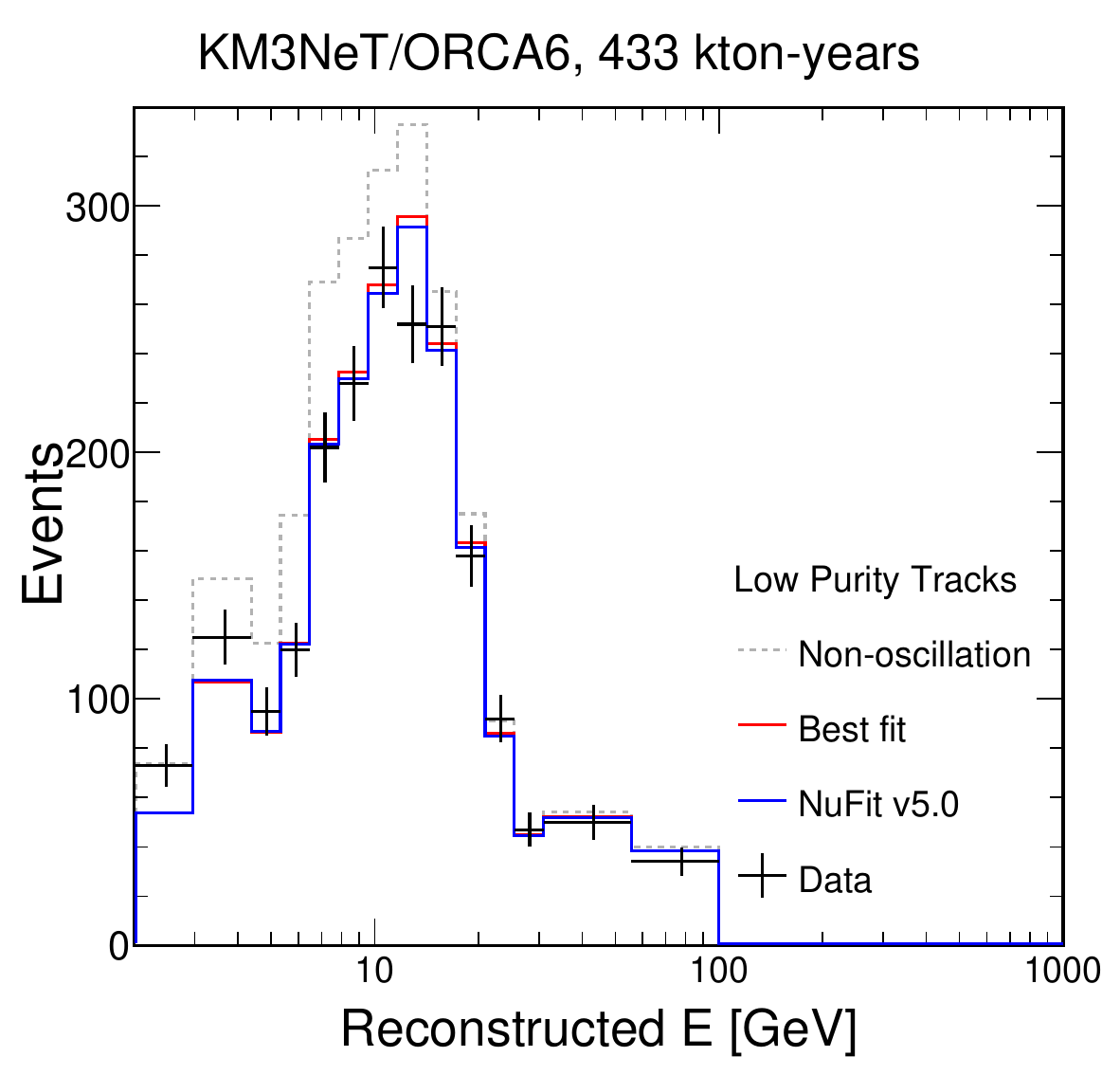}  \includegraphics[width=0.48\linewidth]{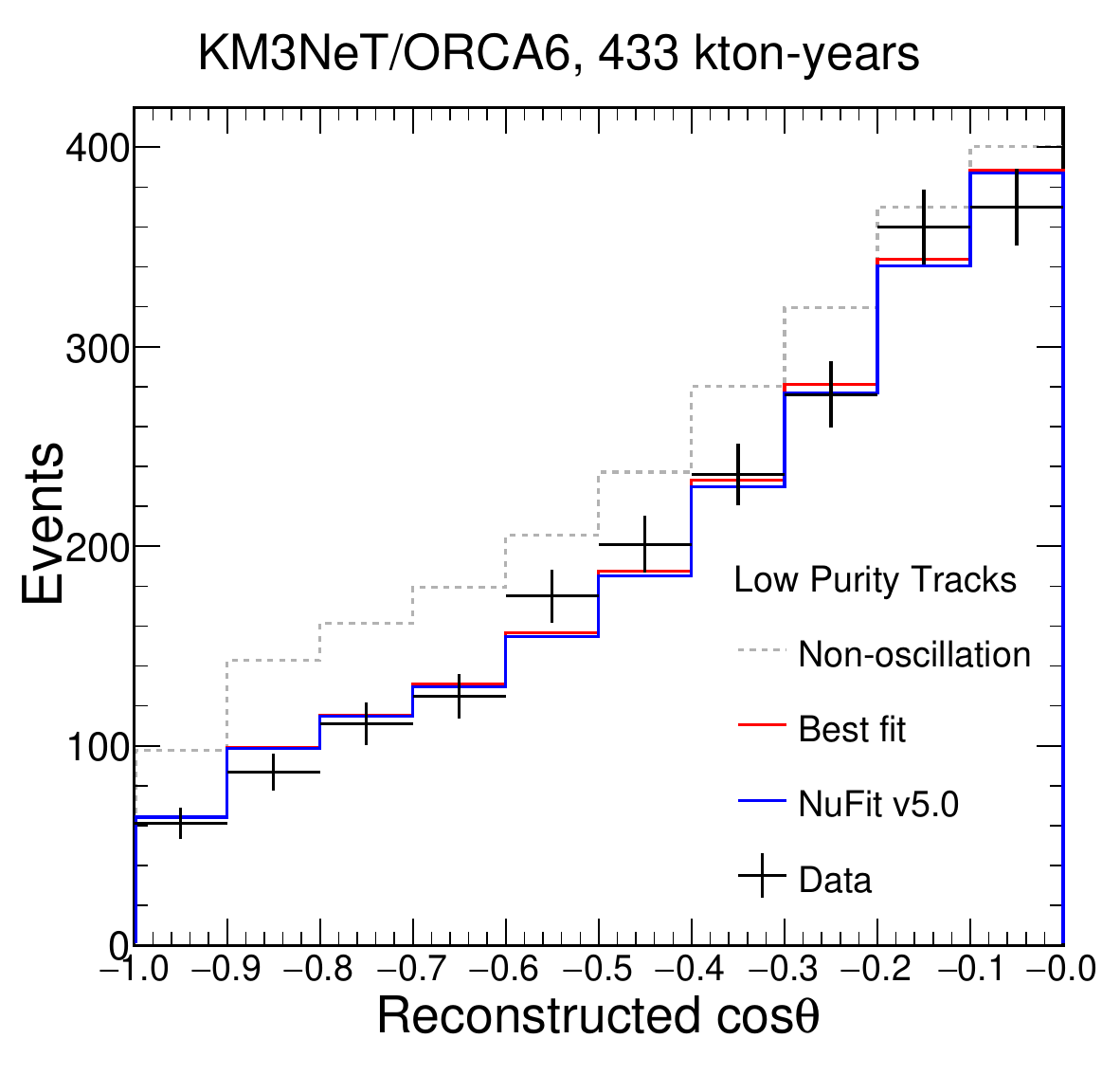} \\
     \includegraphics[width=0.48\linewidth]{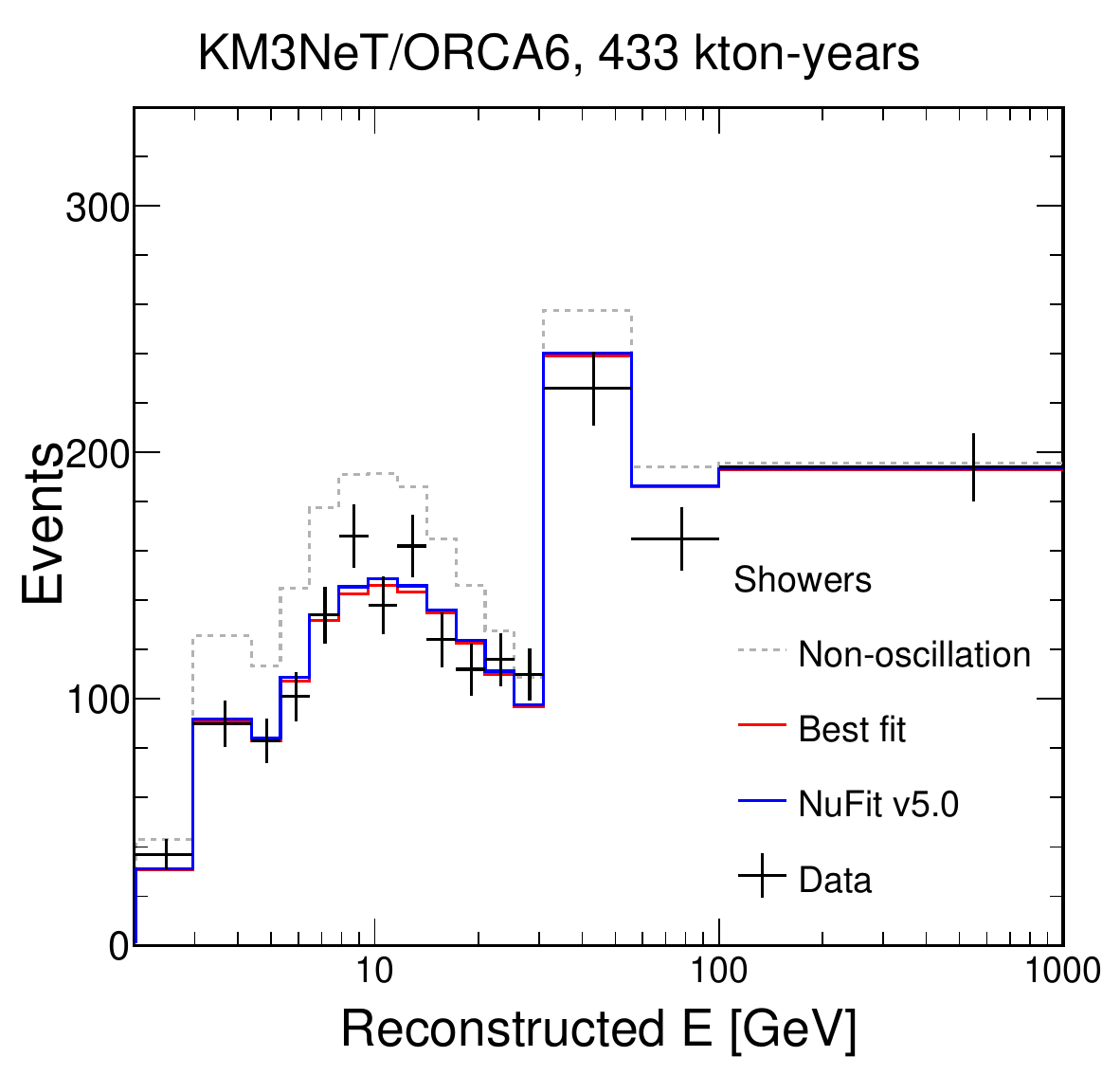}\includegraphics[width=0.48\linewidth]{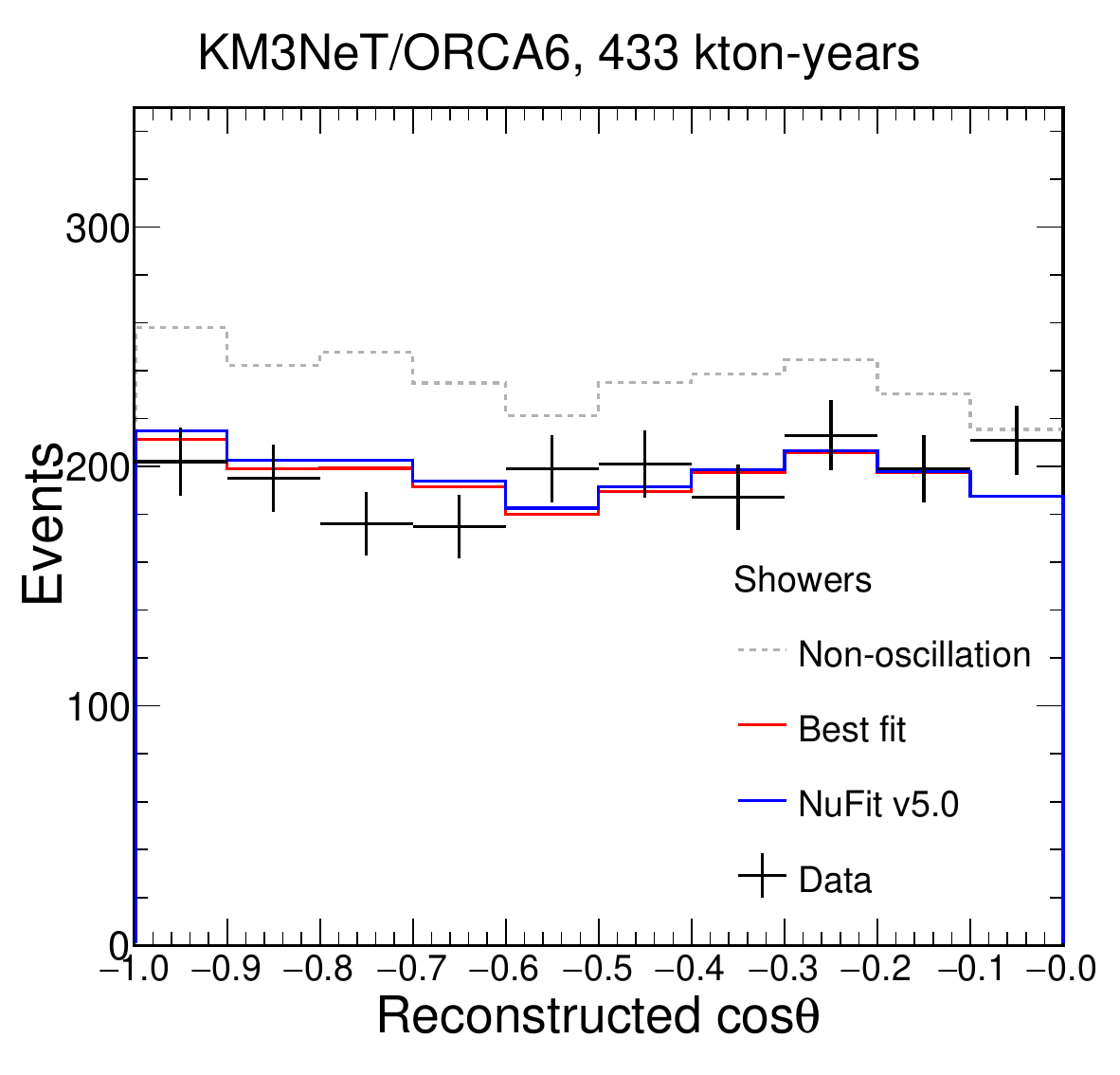}
 \caption{Comparison of event counts as a function of the reconstructed energy (left) and the reconstructed zenith angle (right) for High Purity Tracks (top), Low Purity Tracks (middle) and Showers (bottom). The data points are shown with error bars in black, the best fit is shown in solid red and the oscillation hypothesis with NuFit v5.0 values is shown in solid blue. The non-oscillation hypothesis is indicated as a dashed line.}
 \label{c6-DistrE}
\end{figure}

To visualise the effects of neutrino oscillations, the reconstructed $L/E$ ratio to no-oscillations and the event distributions as functions of reconstructed energy and the  reconstructed cosine of the zenith angle are displayed in figure~\ref{c6-Distr} and figure~\ref{c6-DistrE} respectively. These figures compare the no-oscillation hypothesis, the oscillation fit (best-fit), the NuFit v5.0 hypothesis, and the observed data.  For each hypothesis, the nuisance parameters are set to the ones of the best fit, and only oscillation parameters change. While the High Purity Track and Shower classes are homogeneous in direction, the Low Purity Track class has a majority of horizontal events, because of the higher contamination of atmospheric muons, in particular in the region of horizontal events. The effect of neutrino oscillations is clearly visible in all three classes and is more pronounced for the High Purity Track class. The admixture of neutrino flavours in the Shower sample leads to a more smeared pattern. The underfluctuation in the oscillation dip of the High Purity Tracks for the L/E distribution explains why the constraints on the mixing angle come out better than the expected sensitivity. 

\clearpage
\subsection{Uncertainties in the nuisance parameters}
\label{sec:assym}
The best-fit results and the systematic parameters along with their post-fit uncertainties at 68\% CL are shown in table \ref{MINOS_DOscFree} assuming Wilks' theorem. The post-fit uncertainties are computed by profiling each of them as parameters of interest.

\begin{table}[]
    \centering
\begin{tabular}{|c|c|c|c|}
 \hline
   Parameter & Central value $\pm$ uncertainty& Best Fit & Post-fit uncertainty   \\ \hline
   $f_{\text{all}}$ & 1.00& 1.11& $-$0.14/+0.13 \\ \hline
   $f_{\text{HPT}}$ & 1.00& 0.92& $-$0.04/+0.04 \\ \hline
   $f_{\text{S}}$ & 1.00& 0.92& $-$0.06/+0.06 \\ \hline
   $f_{\text{HE}}$ & $1.00 \pm 0.50$ & 1.59& $-$0.29/+0.32 \\ \hline
   $f_{\mu}$ & 1.00& 0.51& $-$0.35/+0.4 \\ \hline
   $f_{\tau\text{CC}}$ & 1.00 $\pm$ 0.20& 0.92& $-$0.19/+0.19 \\ \hline
   $f_{\text{NC}}$ & 1.00 $\pm$ 0.20 & 0.86& $-$0.19/+0.20 \\ \hline
   $s_{\mu \bar{\mu}}$ & 0.00 $\pm$ 0.05 & 0.00& $-$0.05/+0.05 \\ \hline
   $s_{e \bar{e}}$ & 0.00 $\pm$ 0.07 & 0.01& $-$0.07/+0.07 \\ \hline
   $s_{\mu e}$ & 0.000 $\pm$ 0.020 & $-$0.004& $-$0.020/+0.020 \\ \hline
$\delta_{\gamma}$ & 0.0 $\pm$ 0.3 & $-$0.019& $-$0.025/+0.026 \\ \hline
   $\delta_{\theta}$ & 0.000 $\pm$ 0.020 & $-$0.005& $-$0.019/+0.019 \\ \hline 
   $E_s$ & 1.00 $\pm$ 0.09 & 1.03& $-$0.11/+0.08 \\ 
\hline

\end{tabular}
    \caption{Best-fit values and post-fit uncertainties at 68\% CL of the nuisance parameters from the fit of ORCA6 data to $\Delta m^{2}_{31}$ and $\theta_{23}$. The central value and uncertainty of the Gaussian constrains are reported in the second column.}
       \label{MINOS_DOscFree}
\end{table}

\label{sec:pulls}

To evaluate the impact of the nuisance parameters on the estimation of the parameters of interest the following procedure is performed. For each nuisance parameter, the fit is repeated shifting each value up and down by its own post-fit uncertainty. Then, the overall best-fit value of the parameter of interest is compared with the fit obtained with the "shifted" values.  

\begin{figure}[H]
 \centering
     \includegraphics[height=7cm]{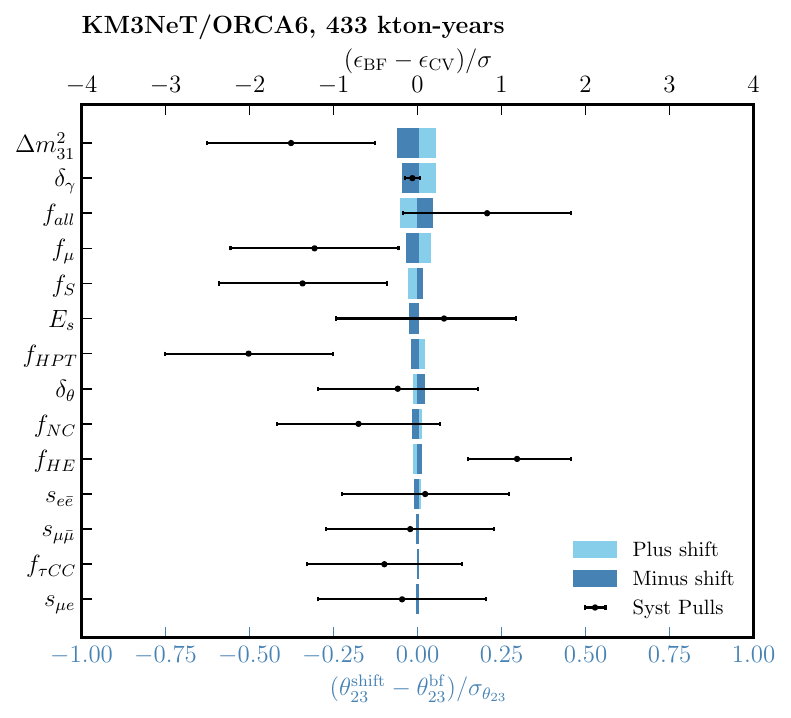}\includegraphics[height=7cm]{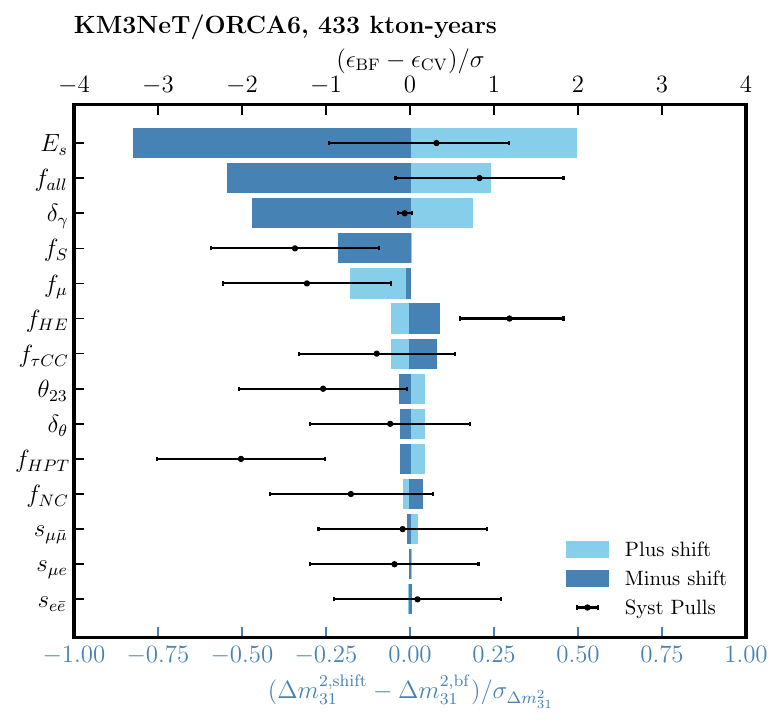}
 
 \caption{Impact of the nuisance parameters on $\theta_{23}$ (left) and $\Delta m^2_{31}$ (right) evaluated repeating the fit shifting the nuisance parameters by their post-fit uncertainties and comparing the fitted value to the best-fit value  (bottom axis). The pulls of the nuisance parameters are reported as dots (top axis) with bars representing the ratio between the post-fit uncertainties and the pre-fit uncertainties (the ratio is set to 1 for unconstrained nuisance parameters). }
 \label{Fig:Star}

\end{figure}

The difference between the nominal best fit of the parameter of interest and the "shifted" value normalised to its $1\sigma$ uncertainty is reported with boxes in figure~\ref{Fig:Star} for each oscillation parameter individually, with the other parameter of interest considered as a nuisance parameter. Additionally, the pulls of the best-fit nuisance parameter values with respect to the central values, ($\epsilon_{\text{BF}}-\epsilon_{\text{CV}})/\sigma$, are reported as dots with error bars where $\sigma$ represents their pre-fit uncertainty. If no pre-fit constraints are given, the post-fit uncertainty is used.  Error bars for the pulls are defined as the ratio between post-fit and pre-fit uncertainties. The error bars of the unconstrained parameters are set to 1.

The spectral index, $\delta_{\gamma}$ and the normalisation of high-energy simulated events, $f_{\text{HE}}$ can be constrained better with the data than with the auxiliary measurements, as can be inferred from the small bars in figure~\ref{Fig:Star}. No significant pulls have been observed in the neutrino flux. The pull in the normalisation of high-energy simulated events is expected as the light simulation software used for high energies leads to a reduced number of selected events with respect to the simulation made with Geant4.

The measurement of the mixing angle is primarily limited by statistical uncertainties, with only a minor contribution from nuisance parameters, particularly the spectral index and overall normalisation. Adjusting and fixing these nuisance parameters results in only a small shift in the mixing angle, indicating a weak dependence. In contrast, the measurement of 
$\Delta m^2_{31}$ is dominated by systematic uncertainties. Shifting and fixing the energy scale $E_s$, overall normalisation $f_{\text{all}}$, or spectral index $\delta_{\gamma}$ significantly alters the fitted value, demonstrating strong correlations with these parameters. While the overall normalisation and spectral index are reasonably well-constrained by the current data and will be more with increased exposure, accurately constraining the detector’s energy response remains a challenge. Substantially more data will be needed to improve this. Additionally, more precise external measurements of the detector’s absolute energy scale are required to further reduce systematic uncertainties, with dedicated studies on PMT efficiencies and water properties being crucial. For more information on the impact of the energy scale on the measurement, see section 7.2 of ref.~\cite{chau:tel-03999509}.

\clearpage

\section{Conclusions}

In a sample of 5828 neutrino candidates corresponding to an exposure of 433 kton-years, the atmospheric neutrino oscillation parameters were measured with the ORCA6 detector:
\begin{align*}
&\sin^2\theta_{23}= 0.51^{+0.04}_{-0.05} \\
     &\Delta m^2_{31} =
    \begin{cases}
      2.18^{+0.25}_{-0.35}\times 10^{-3}~\mathrm{eV^2},\quad \text{for NO} \\
      [-2.25,-1.76]\times 10^{-3}~\mathrm{eV^2}, \quad \text{for IO}
    \end{cases} \\
&2\ln(L_{\text{NO}}/L_{\text{IO}}) = 0.31
\end{align*}
\noindent The IO is disfavoured  with a p-value of 0.25. The errors were computed via the Feldman-Cousins method for 68\% CL and incorporate both the statistical and systematic uncertainties. 

These results indicate that significant contributions to the field of neutrino oscillations are made with the KM3NeT/ORCA detector, even in its early stages with a limited configuration and statistically small sample. The constraints on the oscillation parameters are becoming competitive and align with those obtained by other experiments.

The primary limitations of this study are the poor energy resolution of tracks shown in figure~\ref{Fig:RecoE} and the challenges in track/shower separation at low energies shown in figure~\ref{Fig:C6-PIDdata} (right). These issues are expected to improve as more detection units are installed. Additionally, the uncertainty on the measurement of the mixing angle $\theta_{23}$ is statistically dominated, and will improve significantly  with an increase of the data sample. On the other hand, the uncertainty on the measurement of $\Delta m^2_{31}$ is dominated by the systematic uncertainty on the absolute energy scale of the detector, which relates to the PMT efficiencies and water properties. Dedicated studies on these topics are under way to further constrain these parameters as the detector grows.

\section{Acknowledgements}
\noindent The authors acknowledge the financial support of the funding agencies:
%Belgium
Funds for Scientific Research (FRS-FNRS), Francqui foundation, BAEF foundation.
%Czech
Czech Science Foundation (GAČR 24-12702S);
%France
Agence Nationale de la Recherche (contract ANR-15-CE31-0020), Centre National de la Recherche Scientifique (CNRS), Commission Europ\'eenne (FEDER fund and Marie Curie Program), LabEx UnivEarthS (ANR-10-LABX-0023 and ANR-18-IDEX-0001), Paris \^Ile-de-France Region, Normandy Region (Alpha, Blue-waves and Neptune), France;
%Georgia
Shota Rustaveli National Science Foundation of Georgia (SRNSFG, FR-22-13708), Georgia;
%Germany (Max Planck Inst.)
This work is part of the MuSES project which has received funding from the European Research Council (ERC) under the European Union’s Horizon 2020 Research and Innovation Programme (grant agreement No 101142396).
%Greece
The General Secretariat of Research and Innovation (GSRI), Greece;
%Italy
Istituto Nazionale di Fisica Nucleare (INFN) and Ministero dell’Universit{\`a} e della Ricerca (MUR), through PRIN 2022 program (Grant PANTHEON 2022E2J4RK, Next Generation EU) and PON R\&I program (Avviso n. 424 del 28 febbraio 2018, Progetto PACK-PIR01 00021), Italy; IDMAR project Po-Fesr Sicilian Region az. 1.5.1; A. De Benedittis, W. Idrissi Ibnsalih, M. Bendahman, A. Nayerhoda, G. Papalashvili, I. C. Rea, A. Simonelli have been supported by the Italian Ministero dell'Universit{\`a} e della Ricerca (MUR), Progetto CIR01 00021 (Avviso n. 2595 del 24 dicembre 2019); KM3NeT4RR MUR Project National Recovery and Resilience Plan (NRRP), Mission 4 Component 2 Investment 3.1, Funded by the European Union – NextGenerationEU,CUP I57G21000040001, Concession Decree MUR No. n. Prot. 123 del 21/06/2022;
%Morocco
Ministry of Higher Education, Scientific Research and Innovation, Morocco, and the Arab Fund for Economic and Social Development, Kuwait;
%The Netherlands
Nederlandse organisatie voor Wetenschappelijk Onderzoek (NWO), the Netherlands;
%Romania
Ministry of Research, Innovation and Digitalisation, Romania;
%Slovak Republic
Slovak Research and Development Agency under Contract No. APVV-22-0413; Ministry of Education, Research, Development and Youth of the Slovak Republic;
%Spain
MCIN for PID2021-124591NB-C41, -C42, -C43 and PDC2023-145913-I00 funded by MCIN/AEI/10.13039/501100011033 and by “ERDF A way of making Europe”, for ASFAE/2022/014 and ASFAE/2022 /023 with funding from the EU NextGenerationEU (PRTR-C17.I01) and Generalitat Valenciana, for Grant AST22\_6.2 with funding from Consejer\'{\i}a de Universidad, Investigaci\'on e Innovaci\'on and Gobierno de Espa\~na and European Union - NextGenerationEU, for CSIC-INFRA23013 and for CNS2023-144099, Generalitat Valenciana for CIDEGENT/2018/034, /2019/043, /2020/049, /2021/23, for CIDEIG/2023/20, for CIPROM/2023/51 and for GRISOLIAP/2021/192 and EU for MSC/101025085, Spain;
%UAE
Khalifa University internal grants (ESIG-2023-008 and RIG-2023-070), United Arab Emirates;
%UK
The European Union's Horizon 2020 Research and Innovation Programme (ChETEC-INFRA - Project no. 101008324).

\newpage
\bibliographystyle{JHEP}
\bibliography{main}

\providecommand{\href}[2]{#2}\begingroup\raggedright\begin{thebibliography}{10}

\bibitem{RevModPhys.88.030501}
T.~Kajita, \emph{Nobel Lecture: Discovery of atmospheric neutrino oscillations}, \href{https://doi.org/10.1103/RevModPhys.88.030501}{\emph{Rev. Mod. Phys.} {\bfseries 88} (2016) 030501}.

\bibitem{Pontecorvo1957MesoniumAA}
B.~Pontecorvo, \emph{{Mesonium and anti-mesonium}}, {\emph{\href{http://jetp.ras.ru/cgi-bin/dn/e_006_02_0429.pdf}{Sov. Phys. JETP \textnormal{6 (1957) 429}}\hspace{-3pt}} }.

\bibitem{Maki:1962mu}
Z.~Maki, M.~Nakagawa and S.~Sakata, \emph{{Remarks on the Unified Model of Elementary Particles}}, \href{https://doi.org/10.1143/PTP.28.870}{\emph{PTP} {\bfseries 28} (1962) 870}.

\bibitem{Esteban:2020cvm}
I.~Esteban, M.C.~Gonzalez-Garcia, M.~Maltoni, T.~Schwetz and A.~Zhou, \emph{{The fate of hints: updated global analysis of three-flavor neutrino oscillations}}, \href{https://doi.org/10.1007/JHEP09(2020)178}{\emph{JHEP} {\bfseries 09} (2020) 178}.

\bibitem{deSalas:2020pgw}
P.F.~de~Salas, D.V.~Forero, S.~Gariazzo, P.~Mart\'\i{}nez-Mirav\'e, O.~Mena, C.A.~Ternes et~al., \emph{{2020 global reassessment of the neutrino oscillation picture}}, \href{https://doi.org/10.1007/JHEP02(2021)071}{\emph{JHEP} {\bfseries 02} (2021) 071}.

\bibitem{Capozzi:2021fjo}
F.~Capozzi, E.~Di~Valentino, E.~Lisi, A.~Marrone, A.~Melchiorri and A.~Palazzo, \emph{{Unfinished fabric of the three neutrino paradigm}}, \href{https://doi.org/10.1103/PhysRevD.104.083031}{\emph{Phys. Rev. D} {\bfseries 104} (2021) 083031}.

\bibitem{joao_coelho_2023_10104847}
J.~Coelho, R.~Pestes, A.~Domi, S.~Bourret, U.~Rahaman, L.~Maderer et~al., \emph{joaoabcoelho/OscProb: v2.0.12},  Nov., 2023, \href{https://doi.org/10.5281/zenodo.10104847}{10.5281/zenodo.10104847}.

\bibitem{PREM}
A.M.~Dziewonski and D.L.~Anderson, \emph{Preliminary reference Earth model}, \href{https://doi.org/https://doi.org/10.1016/0031-9201(81)90046-7}{\emph{PEPI} {\bfseries 25} (1981) 297}.

\bibitem{Adri_n_Mart_nez_2016}
{\scshape KM3NeT} collaboration, \emph{Letter of intent for {KM}3NeT 2.0}, \href{https://doi.org/10.1088/0954-3899/43/8/084001}{\emph{J. Phys. G} {\bfseries 43} (2016) 084001}.

\bibitem{Smirnov:2003da}
A.Y.~Smirnov, \emph{{The MSW effect and solar neutrinos}},  in \emph{{10th International Workshop on Neutrino Telescopes}}, pp.~23--43, 5, 2003 [\href{https://arxiv.org/abs/hep-ph/0305106}{{\ttfamily hep-ph/0305106}}].

\bibitem{KM3NeT:2021hkj}
L.~Nauta, \emph{{First neutrino oscillation measurement in KM3NeT/ORCA}}, \href{https://doi.org/10.22323/1.395.1123}{\emph{PoS} {\bfseries ICRC2021} (2021) 1123}.

\bibitem{Carretero:2022mjc}
V.~Carretero, \emph{{Searches for neutrino physics beyond the standard model with KM3NeT/ORCA6}}, \href{https://doi.org/10.22323/1.414.0578}{\emph{PoS} {\bfseries ICHEP2022} (2022) 578}.

\bibitem{ANTARES:1999fhm}
{\scshape ANTARES} collaboration, \emph{ANTARES: The first undersea neutrino telescope}, \href{https://doi.org/https://doi.org/10.1016/j.nima.2011.06.103}{\emph{Nucl. Instrum. Methods Phys. Res.} {\bfseries 656} (2011) 11}.

\bibitem{NMOpaper}
{\scshape KM3NeT} collaboration, \emph{{Determining the neutrino mass ordering and oscillation parameters with KM3NeT/ORCA}}, \href{https://doi.org/10.1140/epjc/s10052-021-09893-0}{\emph{Eur. Phys. J. C} {\bfseries 82} (2022) 26}.

\bibitem{KM3NeT:2021rkn}
{\scshape KM3NeT, JUNO} collaboration, \emph{{Combined sensitivity of JUNO and KM3NeT/ORCA to the neutrino mass ordering}}, \href{https://doi.org/10.1007/JHEP03(2022)055}{\emph{JHEP} {\bfseries 03} (2022) 055}.

\bibitem{KM3NeT:2021uez}
{\scshape KM3NeT} collaboration, \emph{{Sensitivity to light sterile neutrino mixing parameters with KM3NeT/ORCA}}, \href{https://doi.org/10.1007/JHEP10(2021)180}{\emph{JHEP} {\bfseries 10} (2021) 180}.

\bibitem{Decay_ORCA_2023}
{\scshape KM3NeT} collaboration, \emph{Probing invisible neutrino decay with {KM}3NeT/{ORCA}}, \href{https://doi.org/10.1007/jhep04(2023)090}{\emph{JHEP} {\bfseries 2023} (2023) 90}.

\bibitem{Domi:2023uy}
A.~Domi, \emph{{Lorentz Invariance Violation with KM3NeT/ORCA115}}, \href{https://doi.org/10.22323/1.444.1086}{\emph{PoS} {\bfseries ICRC2023} (2023) 1086}.

\bibitem{KM3NeT:2022pnv}
{\scshape KM3NeT} collaboration, \emph{{The KM3NeT multi-PMT optical module}}, \href{https://doi.org/10.1088/1748-0221/17/07/P07038}{\emph{JINST} {\bfseries 17} (2022) P07038}.

\bibitem{75839}
D.~Lefevre, M.~Libes, D.~Mallarino, K.~Bernardet and C.~Gojak, \emph{EMSO-Ligure Ouest observatory data (MII) from 2019-08},  2024.
\newblock \href{https://doi.org/10.17882/75839}{10.17882/75839}.

\bibitem{Aiello_2018}
{\scshape KM3NeT} collaboration, \emph{Characterisation of the Hamamatsu photomultipliers for the KM3NeT Neutrino Telescope}, \href{https://doi.org/10.1088/1748-0221/13/05/P05035}{\emph{J. Instrum.} {\bfseries 13} (2018) P05035}.

\bibitem{annurev-marine-120308-081028}
S.H.~Haddock, M.A.~Moline and J.F.~Case, \emph{Bioluminescence in the Sea}, \href{https://doi.org/10.1146/annurev-marine-120308-081028}{\emph{Annu. Rev. Mar. Sci.} {\bfseries 2} (2010) 443}.

\bibitem{Albert_2018}
{\scshape ANTARES} collaboration, \emph{Long-term monitoring of the ANTARES optical module efficiencies using $^{40}\mathrm{{K} }$ decays in sea water}, \href{https://doi.org/10.1140/epjc/s10052-018-6132-2}{\emph{Eur. Phys. J. C} {\bfseries 78} (2018) }.

\bibitem{Formaggio_2012}
J.A.~Formaggio and G.P.~Zeller, \emph{From eV to EeV: Neutrino cross sections across energy scales}, \href{https://doi.org/10.1103/revmodphys.84.1307}{\emph{Rev. Mod. Phys.} {\bfseries 84} (2012) 13071}.

\bibitem{PDG}
{\scshape PARTICLE DATA GROUP} collaboration, \emph{{Review of Particle Physics}}, \href{https://doi.org/10.1093/ptep/ptac097}{\emph{PTEP} {\bfseries 2022} (2022) 083C01}.

\bibitem{refId0}
T.~Chiarusi, E.~Giorgio and D.~Zito, \emph{The KM3NeT data acquisition system - Status and evolution}, \href{https://doi.org/10.1051/epjconf/202328008004}{\emph{EPJ Web Conf.} {\bfseries 280} (2023) 08004}.

\bibitem{OFearraigh:2024ioy}
B.~\'O~Fearraigh, \emph{{Following the light: Novel event reconstruction techniques for neutrino oscillation analyses in KM3NeT/ORCA}}, Ph.D. thesis, Amsterdam U., 2024, \href{https://www.nikhef.nl/pub/services/biblio/theses\_pdf/thesis\_B\_Ofearraigh.pdf}{https://www.nikhef.nl/pub/services/biblio/theses\_pdf/thesis\_B\_Ofearraigh.pdf}.

\bibitem{Alba}
A.~Domi, \emph{Shower reconstruction and sterile neutrino analysis with KM3NeT/ORCA and ANTARES}, Ph.D. thesis, Universit\'{a} degli studi di Genova, 2018, \href{http://hdl.handle.net/11567/985989}{http://hdl.handle.net/11567/985989}.

\bibitem{KM3NeT:2021lsb}
{\scshape KM3NeT} collaboration, \emph{{Nanobeacon: A time calibration device for the KM3NeT neutrino telescope}}, \href{https://doi.org/10.1016/j.nima.2022.167132}{\emph{Nucl. Instrum. Meth. A} {\bfseries 1040} (2022) 167132}.

\bibitem{Bailly-Salins:2023IA}
L.~Bailly-Salins, \emph{{Time, position and orientation calibration using atmospheric muons in KM3NeT}}, \href{https://doi.org/10.22323/1.444.0218}{\emph{PoS} {\bfseries ICRC2023} (2023) 218}.

\bibitem{SanchezLosa:2023v3}
A.~Sánchez~Losa, J.~Palacios~Gonzalez, F.~Salesa~Greus, J.~Zúñiga~Román, D.~Real~Máñez and D.~Calvo Díaz-Aldagalán, \emph{{KM3NeT Time calibration with Nanobeacons}}, \href{https://doi.org/10.22323/1.444.1062}{\emph{PoS} {\bfseries ICRC2023} (2023) 1062}.

\bibitem{GatiusOliver:2023qr}
C.~Gatius~Oliver, F.~Bretaudeau, M.~De~Jong and L.~Martin, \emph{{Dynamical position and orientation calibration of the KM3NeT telescope}}, \href{https://doi.org/10.22323/1.444.1033}{\emph{PoS} {\bfseries ICRC2023} (2023) 1033}.

\bibitem{Aiello_2020}
{\scshape KM3NeT} collaboration, \emph{gSeaGen: The KM3NeT GENIE-based code for neutrino telescopes}, \href{https://doi.org/10.1016/j.cpc.2020.107477}{\emph{Comput. Phys. Commun.} {\bfseries 256} (2020) 107477}.

\bibitem{andreopoulos2015genie}
C.~Andreopoulos, C.~Barry, S.~Dytman, H.~Gallagher, T.~Golan, R.~Hatcher et~al., \emph{{The GENIE Neutrino Monte Carlo Generator: Physics and User Manual}},  \href{https://arxiv.org/abs/1510.05494}{{\ttfamily 1510.05494}}.

\bibitem{Tsirigotis:2011zza}
A.~Tsirigotis, A.~Leisos and S.~Tzamarias, \emph{{HOU Reconstruction \& Simulation (HOURS): A complete simulation and reconstruction package for very large volume underwater neutrino telescopes}}, \href{https://doi.org/10.1016/j.nima.2010.06.258}{\emph{Nucl. Instrum. Meth. A} {\bfseries 626-627} (2011) S185}.

\bibitem{AGOSTINELLI2003250}
{\scshape GEANT4} collaboration, \emph{Geant4—a simulation toolkit}, \href{https://doi.org/https://doi.org/10.1016/S0168-9002(03)01368-8}{\emph{Nucl. Instrum. Meth. A} {\bfseries 506} (2003) 250}.

\bibitem{Carminati:2008qb}
M.~Bazzoti, G.~Carminati, A.~Margiotta and M.~Spurio, \emph{{Atmospheric MUons from PArametric formulas: A Fast GEnerator for neutrino telescopes (MUPAGE)}}, \href{https://doi.org/10.1016/j.cpc.2008.07.014}{\emph{Comput.\ Phys.\ Commun.} {\bfseries 179} (2008) 915}.

\bibitem{Becherini:2005sr}
Y.~Becherini, A.~Margiotta, M.~Sioli and M.~Spurio, \emph{{A Parameterisation of single and multiple muons in the deep water or ice}}, \href{https://doi.org/10.1016/j.astropartphys.2005.10.005}{\emph{Astropart. Phys.} {\bfseries 25} (2006) 1}.

\bibitem{Chen_2016}
T.~Chen and C.~Guestrin, \emph{XGBoost: A Scalable Tree Boosting System},  in \emph{Proceedings of the 22nd ACM SIGKDD International Conference on Knowledge Discovery and Data Mining}, KDD '16, (New York, NY, USA), p.~785–794, Association for Computing Machinery, 2016, \href{https://doi.org/10.1145/2939672.2939785}{https://doi.org/10.1145/2939672.2939785}.

\bibitem{Bourret:2018kug}
S.~Bourret, \emph{{Neutrino oscillations and earth tomography with KM3NeT-ORCA}}, Ph.D. thesis, APC, Paris, 2018, \href{https://tel.archives-ouvertes.fr/tel-02491394/file/BOURRET\_Simon\_1\_complete\_20181130.pdf}{https://tel.archives-ouvertes.fr/tel-02491394/file/BOURRET\_Simon\_1\_complete\_20181130.pdf}.

\bibitem{HondaFlux}
M.~Honda, M.S.~Athar, T.~Kajita, K.~Kasahara and S.~Midorikawa, \emph{Atmospheric neutrino flux calculation using the NRLMSISE-00 atmospheric model}, \href{https://doi.org/10.1103/PhysRevD.92.023004}{\emph{Phys. Rev. D} {\bfseries 92} (2015) 023004}.

\bibitem{Cousins2013GeneralizationOC}
R.D.~Cousins, \emph{Generalization of Chisquare Goodness-ofFit Test for Binned Data Using Saturated Models , with Application to Histograms},  2013, \href{https://api.semanticscholar.org/CorpusID:5936965}{https://api.semanticscholar.org/CorpusID:5936965}.

\bibitem{BAKER1984437}
S.~Baker and R.D.~Cousins, \emph{{Clarification of the use of chi square and likelihood functions in fits to histograms}}, \href{https://doi.org/10.1016/0167-5087(84)90016-4}{\emph{Nucl. Instrum. Meth.} {\bfseries 221} (1984) 437}.

\bibitem{BARLOW1993219}
R.~Barlow and C.~Beeston, \emph{Fitting using finite Monte Carlo samples}, \href{https://doi.org/https://doi.org/10.1016/0010-4655(93)90005-W}{\emph{Comput.\ Phys.\ Commun.} {\bfseries 77} (1993) 219}.

\bibitem{Conway:2011in}
J.S.~Conway, \emph{{Incorporating Nuisance Parameters in Likelihoods for Multisource Spectra}},  in \emph{{PHYSTAT 2011}}, pp.~115--120, 2011 [\href{https://arxiv.org/abs/1103.0354}{{\ttfamily 1103.0354}}].

\bibitem{PhysRevD.74.094009}
G.D.~Barr, S.~Robbins, T.K.~Gaisser and T.~Stanev, \emph{Uncertainties in atmospheric neutrino fluxes}, \href{https://doi.org/10.1103/PhysRevD.74.094009}{\emph{Phys. Rev. D} {\bfseries 74} (2006) 094009}.

\bibitem{Evans_2017}
J.~Evans, D.~Garcia~Gamez, S.D.~Porzio, S.~S\"oldner-Rembold and S.~Wren, \emph{Uncertainties in atmospheric muon-neutrino fluxes arising from cosmic-ray primaries}, \href{https://doi.org/10.1103/PhysRevD.95.023012}{\emph{Phys. Rev. D} {\bfseries 95} (2017) 023012}.

\bibitem{chau:tel-03999509}
T.N.~Chau, \emph{{Study of atmospheric neutrino oscillations with the deep-sea Cherenkov detector KM3NeT/ORCA and synergies with reactor neutrinos}}, Ph.D. thesis, {Universit{\'e} Paris Cit{\'e}}, Nov., 2021, \href{https://theses.hal.science/tel-03999509/file/va\_Chau\_Thien\_Nhan.pdf}{https://theses.hal.science/tel-03999509/file/va\_Chau\_Thien\_Nhan.pdf}.

\bibitem{Cranmer:2014lly}
K.~Cranmer, \emph{{Practical Statistics for the LHC}},  in \emph{{2011 European School of High-Energy Physics}}, pp.~267--308, 2014, \href{http://dx.doi.org/10.5170/CERN-2014-003.267}{http://dx.doi.org/10.5170/CERN-2014-003.267}.

\bibitem{Feldman_1998}
G.J.~Feldman and R.D.~Cousins, \emph{Unified approach to the classical statistical analysis of small signals}, \href{https://doi.org/10.1103/physrevd.57.3873}{\emph{Phys. Rev. D} {\bfseries 57} (1998) 3873}.

\bibitem{s__s__wilks_1938}
S.S.~Wilks, \emph{The Large-Sample Distribution of the Likelihood Ratio for Testing Composite Hypotheses}, \href{https://doi.org/10.1214/AOMS/1177732360}{\emph{Annals of Mathematical Statistics} {\bfseries 9} (1938) 60}.

\bibitem{icecubecollaboration2024measurement}
{\scshape IceCube} collaboration, \emph{Measurement of atmospheric neutrino oscillation parameters using convolutional neural networks with 9.3 years of data in IceCube DeepCore},  \href{https://arxiv.org/abs/2405.02163}{{\ttfamily 2405.02163}}.

\bibitem{abe2023updated}
{\scshape T2K} collaboration, \emph{{Updated T2K measurements of muon neutrino and antineutrino disappearance using 3.6\texttimes{}1021 protons on target}}, \href{https://doi.org/10.1103/PhysRevD.108.072011}{\emph{Phys. Rev. D} {\bfseries 108} (2023) 072011}.

\bibitem{yasuhiro_nakajima_2020_4134680}
Y.~Takeuchi, \emph{Recent results and future prospects of Super-Kamiokande}, \href{https://doi.org/https://doi.org/10.1016/j.nima.2018.11.093}{\emph{Nucl. Instrum. Methods Phys. Res.} {\bfseries 952} (2020) 161634}.

\bibitem{aurisano_2018_1286760}
A.~{Aurisano}, \emph{{Recent Results from MINOS and MINOS+}},  in \emph{XXVIII International Conference on Neutrino Physics and Astrophysics}, p.~423, June, 2018, \href{https://doi.org/10.5281/zenodo.1286760}{10.5281/zenodo.1286760}.

\bibitem{alex_himmel_2020_4142045}
A.~Himmel, \emph{New Oscillation Results from the NOvA Experiment},  Oct., 2020, \href{https://doi.org/10.5281/zenodo.4142045}{10.5281/zenodo.4142045}.

\bibitem{article}
S.~Algeri, J.~Aalbers, K.~Morå and J.~Conrad, \emph{Searching for new phenomena with profile likelihood ratio tests}, \href{https://doi.org/10.1038/s42254-020-0169-5}{\emph{Nature Reviews Physics} {\bfseries 2} (2020) 1}.

\end{thebibliography}\endgroup
\end{document}